\documentclass[review]{elsarticle}
\usepackage{amsmath,amssymb}
\usepackage{bbm}
\usepackage{graphicx,graphics}
\usepackage{lineno,hyperref}
\usepackage{soul}
\usepackage[makeroom]{cancel}
\usepackage{comment}
\usepackage[normalem]{ulem} 
\usepackage{xcolor} 

\modulolinenumbers[1]

\journal{Applied Energy}

\begin{document}

\begin{frontmatter}

\title{Super-relaxation of space-time-quantized ensemble of energy loads to curtail their synchronization after demand response perturbation}

\author[mymainaddress1,mysecondaryaddress]{I. Luchnikov}

\author[mymainaddress2,mysecondaryaddress2]{D. M\'etivier}

\author[mymainaddress1]{H. Ouerdane}
\cortext[mycorrespondingauthor]{Corresponding author}
\ead{h.ouerdane@skoltech.ru}

\author[mymainaddress3,mymainaddress1]{M. Chertkov}

\address[mymainaddress1]{Skolkovo Institute of Science and Technology, Moscow 121205, Russia}
\address[mysecondaryaddress]{{Moscow Institute of Physics and Technology, Dolgoprudny, Moscow Region 141700, Russia}}
\address[mymainaddress2]{Theoretical Division, Los Alamos National Laboratory, Los Alamos, New Mexico 87545, USA}
\address[mysecondaryaddress2]{Centre de Mathématiques Appliquées (CMAP), Ecole Polytechnique, Route de Saclay, 91128 Palaiseau Cedex, France}
\address[mymainaddress3]{Program in Applied Mathematics, University of Arizona, Tucson, AZ  85721, USA}

\begin{abstract}
Ensembles of thermostatically controlled loads (TCL) provide a significant demand response reserve for the system operator to balance power grids. However, this also results in the parasitic synchronization of individual devices within the ensemble leading to long post-demand-response oscillations in the integrated energy consumption of the ensemble. The synchronization is eventually destructed by fluctuations, thus leading to the (pre-demand response) steady state; however, this natural desynchronization, or relaxation to a statistically steady state, is too long. A resolution of this problem consists in measuring the ensemble's instantaneous consumption and using it as a feedback to stochastic switching of the ensemble's devices between on- and off- states. A simplified continuous-time model showed that carefully tuned nonlinear feedback results in a fast (super-) relaxation of the ensemble energy consumption. Since both state information and control signals are discrete, the actual TCL devices operation is space-time quantized, and this must be considered for realistic TCL ensemble modelling. Here, assuming that states are characterized by indoor temperature (quantifying comfort) and air conditioner regime (on, off), we construct a discrete model based on the probabilistic description of state transitions. We demonstrate that super-relaxation holds in such a more realistic setting, and that while it is stable against randomness in the stochastic matrix of the quantized model, it remains sensitive to the time discretization scheme. Aiming to achieve a balance between super-relaxation and customer's comfort, we analyze the dependence of super-relaxation on details of the space-time quantization, and provide a simple analytical criterion to avoid undesirable oscillations in consumption.
\end{abstract}

\begin{keyword}
Demand response; thermostatically controlled loads; energy consumption dynamics
\end{keyword}

\end{frontmatter}


\section{Introduction}
Power grids of today are uncertain with the major sources of uncertainty being fluctuations of renewables, especially of wind and solar, and market uncertainty. To deal with the uncertainties, grid operators need new flexible and inexpensive resources. Demand response (DR) came up prominently as a way if not to resolve the problem completely, then at least to reduce its consequences \cite{OConnell2014}. The main idea consists in exploiting the fact that many consumers of electricity, also called loads, can tolerate delays provided that their comfort zone is not violated.  

The complexity of the power system and the electricity markets dynamics call for the development of practical approaches capable to implement DR while accounting for the increasing penetration of the renewables. Demand side management and automatic meter management systems on the customers' side, and on the supply side remote control of power flows, have been proposed to facilitate renewables' integration \cite{Hammons2008}. To mitigate risks associated with imbalances of the supply and demand, balancing demand response on an hourly basis was suggested \cite{Critz2013}. A market-based approach with correct price signals, fostering flexible, smart power system has been discussed as a way forward to integrate large shares of renewables \cite{Auer2016}. Further, while DR presents benefits and costs \cite{Albadi2007}, which depend on the technology used for control schemes \cite{Lampropoulos2013}, it can reduce the market prices and consumers' cost owing to its contribution to the capacity market \cite{Lynch2019}. Moderate energy consumers like the service sector may participate in DR if barriers (e.g., restructuring costs and regulations) are mitigated and drivers (e.g., positive public image) enhanced \cite{Wohlfarth2020}. As power generation affected by fluctuating renewables, may result in transmission line overload, electricity grids' topology must be designed to sustain such overloads \cite{Schiel2017}. Electric vehicles are increasingly suggested as flexible grid components that permit stability \cite{Gajduk2014,Yesilbudak2018} while partaking in DR. But in case of contingency, distributed DR scheduling can help with frequency regulation \cite{Motalleb2016}.

While involving big stable loads, like aluminium smelters, in DR services is a well established practice, there is also great potential in utilizing opportunities in DR which can be offered by many small loads. To unlock this potential, a statistics-based approach as pioneered in, e.g., \cite{Chong1979}, is necessary as shown by a large body of works. Models within this approach quickly evolved to account for characteristics such as lifestyle and weather \cite{Ihara1981}, and propose a methodology for classification of elementary component loads and aggregation \cite{Chong1984}. Using Fokker-Planck equations \cite{Gardiner2004,VanKampen2007}, large aggregates of loads controlled with thermostats such as heaters and air conditioners, were considered given their important impact on the power system dynamics \cite{Malhame1985,Malhame1988}. Interestingly, as the dynamics of electrical heaters and air conditioners can be characterized as an alternating renewal process, it was shown how consumption data can be used to identify electric load models \cite{ElFerik1994}. Considering aggregated thermostatically controlled loads (TCL), a state-queuing model revealed the important influence of load state diversity on the aggregated profile dynamics, as synchronization leads to unwanted peak loads \cite{Lu2004,Lu2005}. While a direct way to achieve peak shaving is by interruption of power delivery to the loads, other, smarter ways propose to control the loads consumption by variation of the temperature set points thus permitting flexibility on fast time scales, and ensemble diversity \cite{Callaway2009,Callaway2011,Bashash2011}. To ensure efficacy of load management to lower TCLs' aggregated power consumption when needed, by feedback control, a model-based feedback control strategy must aim for high accuracy in the characterization of the aggregate dynamics \cite{Perfumo2012}, as control errors may occur with conventional thermostats \cite{Chassin2015}. To mitigate the negative effects of power fluctuations in the distribution networks it was shown how to coordinate multiple TCL groups employing a two-stage optimization model applied in real-time \cite{Wei2018}. Recent works also considered DR as a perturbation of the TCL ensemble dynamics driving it away from its steady state \cite{Chertkov2017,Metivier2019,Metivier2020}. Relaxation after perturbation is of importance to ensure the stability of the power system. This article contributes this later line of work.

Several hurdles must be overcome to make the DR contribution of many small loads meaningful. It is not economically viable to expect a small load, e.g. a thermostatically controlled loadlike air-conditioner or heater, to be engaged in a sophisticated individual control. Instead, aggregation of many small loads would be a preferred solution \cite{Rajabi2017}. In this scheme the aggregator is an authority receiving DR requests from the system operator and broadcasting the same signal to all their consumers.  It is assumed that the consumers obey and perform the requested action, that is switch off or switch on, follows when requested. An unfortunate side effect of all consumers following the same signal is a  parasitic synchronization/oscillations seen long after engagement of the ensemble in the DR \cite{Callaway2009}. Consumer-specific fluctuations will lead, eventually, through mixing to a decay of oscillation (de-synchronization). However, natural mixing is typically weak, thus leading to long transients, delaying availability of the ensemble for the next DR session. As shown in \cite{Chertkov2017}, the randomization of switching, implemented through the broadcast of a Poisson rate of the switch on/off delay, helps to reduce the mixing time while also providing an acceptable ``comfort zone'' guaranteed to loads. Diversity of loads contributing to the ensemble helps to reduce the mixing time even further \cite{Metivier2019}. 

The solution suggested in \cite{Chertkov2017,Metivier2019} did not depend on any knowledge of the current system state (temperature and switch on/off status). The next significant step in improving control of the ensemble was made in \cite{Metivier2020}, where the following question was addressed: is it possible to set up a viable aggregation model that would rely only on receiving instantaneous integrated consumption of the entire ensemble as a feedback? Notice that even though the absence of the individual response of a load makes the problem of organizing the aggregator control harder, the ability to receive one signal, integrated over the entire ensemble, makes the approach desirable from the viewpoint of keeping the consumption of individual loads private. It was shown in \cite{Metivier2020} that the question just posed has an affirmative answer: making nonlinear feedback on the instantaneous integrated consumption of the ensemble allows to accelerate relaxation (de-synchronization) of the integrated consumption to the steady-state. Notice that this approach, coined the ``mean-field'' control in reference to related methods originating from plasma physics, control, management sciences and applied mathematics \cite{2003HCM,2005WBV,2006LL,2006LLb,2006HMC}, has this strong effect, dubbed super-relaxation \cite{Metivier2020}, only on a specially selected expectation over the ensemble's probability distribution (mean instantaneous consumption) while other expectations over instantaneous probability distribution over the ensemble, continue to relax slowly. 

The model in \cite{Metivier2020} assumed continuous temperature variation and time but operations of the actual TCL devices are space-time quantized. In this work we develop a space-time-quantized model of the TCL ensemble, which incorporates mean-field control, that is feedback on instantaneous total consumption, of the switch on/off rates of all the loads of the ensemble. We show that the super-relaxation effect is also observed in the space-time quantized model, better representing the real-world of energy management than the continuous model studied before. We experiment with the model parameters -- the size of space-time quantization steps and degree of the mean-field control nonlinearity in the Poisson switching on/off rates -- to make a recommendation on the choice of parameters achieving a reasonable balance between fast mixing of the ensemble (faster post-DR restoration) and ``comfort zone'' of the consumers.

The article is organized as follows. In Section \ref{sec:formulation}, we derive and describe the basic equations of our space-time quantized model generalizing the space-time continuous model of \cite{Metivier2020}. Section \ref{sec:experiment} is devoted to  discussion of the numerical results and of the insight they provide. Section \ref{sec:conclusions} is reserved for conclusions and discussion of the path forward. Technical details are presented in Appendices.

\section{Problem formulation}
\label{sec:formulation}

\subsection{Space-time continuous TCL model}

We start with an overview of the basic elements of the continuous model \cite{Metivier2020}. Assume that at every moment $t$ each TCL load is characterized by two parameters: (a) consumer's instantaneous temperature $x(t)$ and, (b) binary state, $\gamma(t)=0,1$, characterizing the on or off state of a consumer's thermal (heating or cooling) device. The dynamics of each TCL in the phase space, characterized by the tuple $\sigma\equiv\{x(t), \gamma(t)\}$, can be complex as it depends on various factors such as operating power, desired temperature, outside temperature, as well as on the local level of noise and uncertainty associated with details of the consumer's operation regime (e.g. frequency of the doors or windows opening, traffic through the consumer space, etc). To manage this complexity, we consider the following set of simplifying assumptions (also focusing without loss of generality on air-conditioning, thus cooling, as our enabling example): 
\begin{itemize}
\item[i/] When the device is switched on, the temperature decreases, and the temperature raises when the device is switched off. We assume that the relaxation of $x(t)$ is linear in both switch on and switch off regimes with the $\pm$ relaxation rates equal to each other by the absolute value. Linearity of $x(t)$ is justified in our work by the assumption that the outside temperature $x_+$ and the minimal possible temperature $x_-$ that can be achieved by the air conditioner being constantly in the on position, are both far apart from the indoor temperature perceived as comfortable by the consumers.

\item[ii/] All TCL devices and their settings are identical, both in terms of their relaxation rates and the temperature extent of the comfort zone. 

\item[iii/] Stochastic effects, associated with device-specific uncertainties, are assumed small and are thus neglected. 

\item[iv/] A TCL does not switch on or off immediately after crossing either of the boundaries of the comfort zone. The switching is delayed according to a Poisson process with rate, $r$. It is assumed that the operator broadcasts the same $r$ to all the consumers.
\end{itemize}

Considered within these assumptions, the basic model of the continuous-time TCL dynamics in the $(x,\gamma)$ space is described by the following set of equations \cite{Chong1979,Ihara1981}:
\begin{eqnarray}
    \frac{dx}{dt} = 
    \begin{cases}
        -\nu, \ \gamma= \ \uparrow,\\
        \phantom{-}\nu, \ \gamma= \ \downarrow,
    \end{cases}
    \label{eq1.}
\end{eqnarray}
\begin{eqnarray}
    \gamma(t + dt) = 
    \begin{cases}
        \downarrow, \ {\rm with \ probability} \ rdt \ {\rm otherwise} \ \gamma(t) \ {\rm and} \ x<x_\downarrow ,\\
        \uparrow, \ {\rm with \ probability} \ rdt \ {\rm otherwise} \ \gamma(t) \ {\rm and}\  x>x_\uparrow,
    \end{cases}
    \label{eq2.}
\end{eqnarray}
where $\pm \nu$ are the cooling/heating rates, and $r$ is the constant rate of (Poisson) switching (from on to off and vice versa) defining switching delay after $x(t)$ crosses the threshold temperature, $x_\downarrow$ or $x_\uparrow$. Then, the two-dimensional probability distribution vector, $P(x|t)$, satisfies the following Fokker-Planck equation:
\begin{eqnarray}
    && \left(\partial_t\begin{pmatrix} 1 & 0\\ 0 & 1\end{pmatrix} -{\cal L}\right)P(x|t)=0,\quad P(x|t)\doteq\begin{pmatrix} P_{\uparrow}(x|t)\\ P_{\downarrow}(x|t)\end{pmatrix},\label{eq3.}\\ && {\cal L}\doteq
     \nu\partial_x \begin{pmatrix} 1 & 0\\ 0 & -1\end{pmatrix}
     -r
     \begin{pmatrix} 
         \theta(x_\downarrow -x) & - \theta(x-x_\uparrow)\\ -\theta(x_\downarrow -x) & \theta(x-x_\uparrow)
     \end{pmatrix},
    \label{eq4.}
\end{eqnarray}
where $\theta$ is the Heaviside step function and $P_{\uparrow},P_{\downarrow}$ are the two components of $P(x|t)$, corresponding to the probability distributions for a consumer to be in the switched-on and switched-off states, respectively. This basic model and generalizations were discussed extensively in \cite{Chertkov2017, Metivier2019}. 

To complete the model one needs to describe actions of the aggregator. We assume that the aggregator has instant access to the integrated consumption of the ensemble. This is realistic in the case when all participants of the ensemble are collocated geographically within the power distribution system, i.e., all reside in the same power distribution feeder area. Then we measure the integrated consumption with a physical device sitting at the sub-station connecting the feeder to the rest of the power system. The instantaneous aggregated consumption of the ensemble is $U(t)=\int dx P_\uparrow(x|t)$, where the integral accounts for the number (proportion) of consumers which are switched on at time $t$. Assuming that all consumers are of the same type, e.g. similar flats or houses with a similar set of devices, we switch to dimensionless characteristics where the power consumed by an individual participant of the ensemble is unity. We then assume that the signal representing the Poisson switching on/off rate, $q(t)$, sent by the aggregator to individual consumers is a functional of $U(t)$: $q(t) = C[U(t)]$, and hence acquires a dynamical character. This type of control is called the "mean-field control" \cite{2003HCM,2005WBV,2006LL,2006LLb,2006HMC,Bensoussan2013} because it involves feedback (in choosing the rate $r$) on the global measured quantity, which is instantaneous integrated consumption of the ensemble. A particular form of the Poisson rate dependence on the integrated consumption, $C[U] = r(2U)^{\frac{s}{2}}$, where $r$ is the basic rate introduced in Eq.~\eqref{eq2.}, and the parameter $s$ controls the degree  of nonlinearity, was considered in \cite{Metivier2020}. 

\subsection{Space-time quantized TCL model}

We now proceed with the details of the space-time-quantized version of  the continuous model summarized in Eq.~\eqref{eq3.}. Using the same set of assumptions as in the continuous model, we  bin the temperature range and denote the quantized space states, $\sigma$. Then, we consider transitions from state $\sigma$ to state $\sigma'$ in discrete time. This space-time-quantized description reflects realistic practice of built-in controls of practical (small and inexpensive) TCLs. In the simplest approach with no feedback (no mean-field control) the transition probability matrix, describing probability for a load to transition from the state $\sigma'$ to the state $\sigma$ during the discrete period of time $t$, reads:

\begin{equation}p_{\sigma \sigma'} = p^{(0)}_{\sigma \sigma'} + rp^{\uparrow}_{\sigma \sigma'} + rp^{\downarrow}_{\sigma \sigma'}
\label{eq:p-sum}
\end{equation}

\noindent where $p^{(0)}_{\sigma\sigma'}$ describes the transition from $\sigma'$ to $\sigma$, associated with cyclic evolution as if it would occur exactly at the thresholds (immediately after entering the discomfort zone) and $r p^{\downarrow}_{\sigma\sigma'}$ and $ r p^{\uparrow}_{\sigma\sigma'}$ are corrections due to the Poisson delay in switching between the on and off state. The term $p^{(0)}_{\sigma\sigma'}$ also includes diffusion which is described by random transitions to neighboring nodes with probability $\epsilon$. The physical meaning of $\epsilon$ can be explained as follows. In a typical real-life situation, $x(t)$ fluctuates due to uncertainties, e.g. linked to incidental movement of people inside,  door and windows opening and closing and related. These fluctuations can be simulated by a white noise acting on $x(t)$. The white noise contributes the diffusion term in the Fokker-Planck equation.

\begin{figure}[h]
    \centering
    \includegraphics[width=1.0\textwidth]{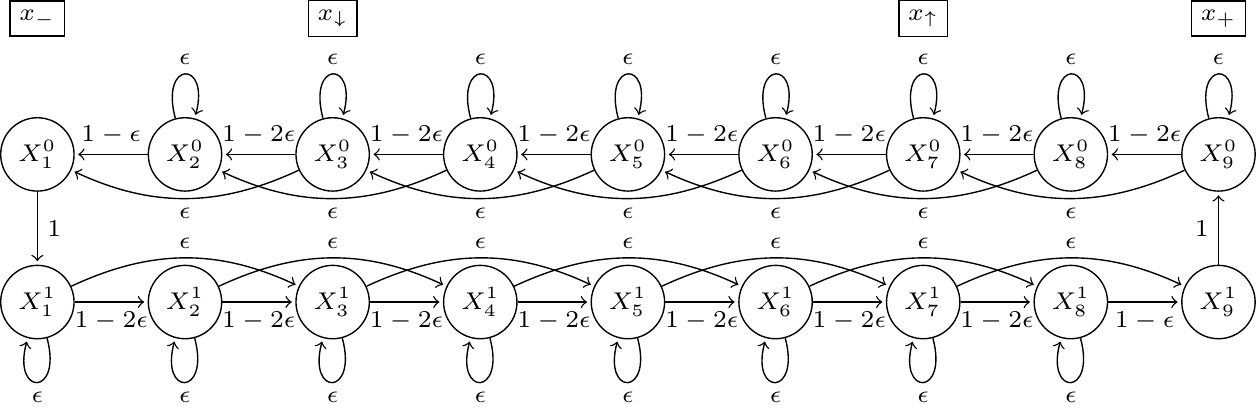}
    \caption{Graphical representation of $p^{(0)}_{\sigma\sigma'}$; $X_i^j$ denotes the state with a particular temperature and regime of the air-conditioner (on, off) where $i$ is the temperature and $j$ is the operation regime of the air-conditioner. The arrows denote the possible transitions with the associated non-zero probability. This part of the transition matrix governs the transitions without Poisson switchings in the out-of-comfort zone. The space between $x_\downarrow$ and $x_\uparrow$ is the comfort zone; $x_-$ and $x_+$ are points where the load must turn to another state. The same-state transitions and the two-step transitions are characterized by a diffusion rate $\epsilon$.}
    \label{fig1}
\end{figure}

\begin{figure}[h]
    \centering
    \includegraphics[width=1.0\textwidth]{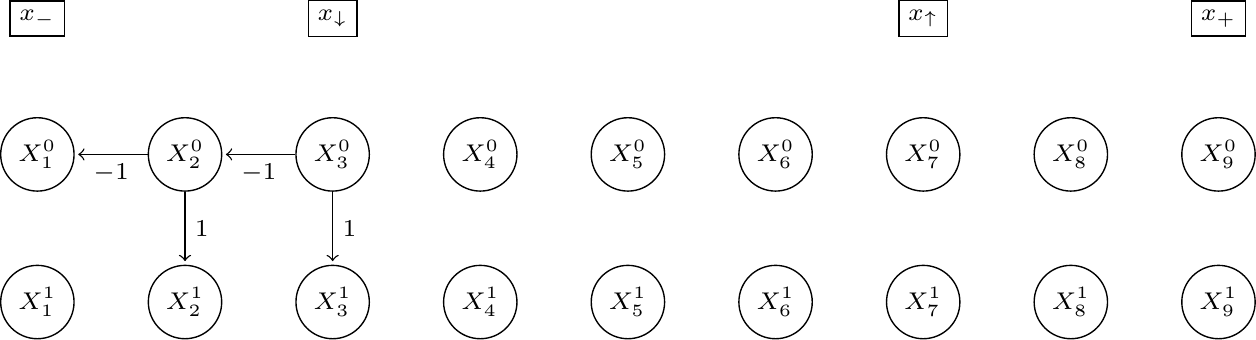}
    \caption{Graphical representation of $p^{\downarrow}_{\sigma \sigma'}$, which governs the Poisson switchings from on to off. Negative elements ensure the preservation of the stochastic property of the resulting transition matrix.}
    \label{fig2}
\end{figure}

\begin{figure}[h]
    \centering
    \includegraphics[width=1.0\textwidth]{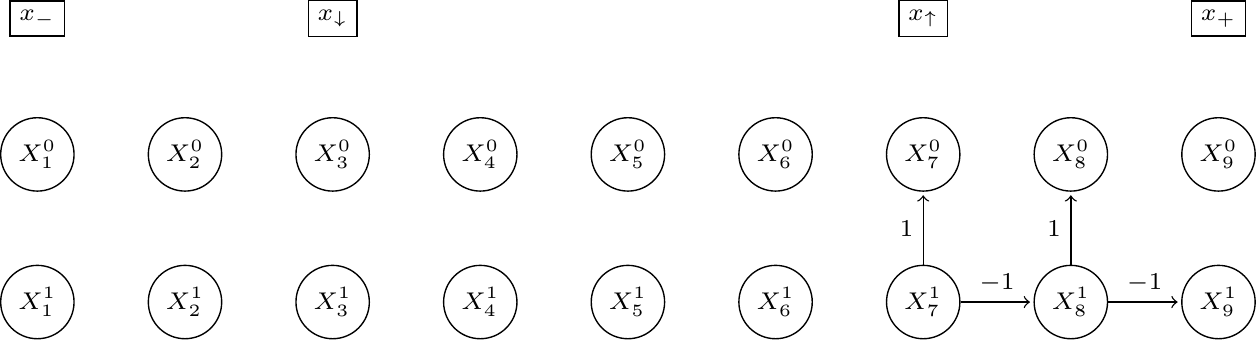}
    \caption{Graphical representation of $p^{\uparrow}_{\sigma \sigma'}$, which governs the Poisson switchings from off to on.}
    \label{fig3}
\end{figure}

The matrices $p_{\sigma\sigma'}^{(0)}$, $p_{\sigma\sigma'}^{\downarrow}$ and $p_{\sigma\sigma'}^{\uparrow}$ can be graphically represented as shown in Figs.~{\ref{fig1}}, {\ref{fig2}}, and {\ref{fig3}} respectively. Notice that the transition probability matrix $p$ defined by Eq.~(\ref{eq:p-sum}) is stochastic, i.e.
\begin{eqnarray}\label{eq:stoch}
\sum_\sigma p_{\sigma \sigma'}=1. 
\end{eqnarray}
Transitions between states are then governed by the following time-space-discrete master equation: 
\begin{equation}
\rho_\sigma(t+1)=\sum_{\sigma'}p_{\sigma\sigma'}\rho_{\sigma'}(t) 
\label{eq:master}
\end{equation}
where $\rho_\sigma(t)$ is a probability mass function, which stands for the probability of a TCL to be in the state $\sigma$ at time $t$.

~\\ 

Mean-field control amounts to allowing the switching rate to be dependent on the energy consumption of a device in a particular state averaged over the probability mass function
\begin{equation}
N_{\uparrow}(t) = \sum_\sigma \rho_\sigma(t) U_\sigma,
\label{up_loads}
\end{equation}

\noindent where 

\begin{equation}
U_\sigma=\begin{cases}1,\ {\rm if}\ \sigma\in {\rm set\ of\ on\ states}\\0, {\rm otherwise}.\end{cases}
\end{equation}

\noindent With this definition, $N_{\uparrow}(t)$ can also be understood as the fraction of loads switched on at time $t$, and we may then generalize Eq.~(\ref{eq:p-sum}) as follows
\begin{eqnarray}
&&p_{\sigma \sigma'}(t)=p^{(0)}_{\sigma\sigma'}+q_{\uparrow}(t)p^{\uparrow}_{\sigma\sigma'}+q_{\downarrow}(t) p^{\downarrow}_{\sigma\sigma'},\nonumber\\
&&q_{\downarrow}(t) = f\left(r[2N_{\uparrow}(t)]^\alpha\right),\nonumber\\
&&q_{\uparrow}(t) = f\left(r[2(1 - N_{\uparrow}(t))]^\alpha\right),
\label{master_eq}
\end{eqnarray}
where $q_{\uparrow}(t)$ and $q_{\downarrow}(t)$ are the Poisson rates modified by the mean-field control, \emph{via} the function $f$ explicitly defined further below, and $\alpha$ denotes the degree of nonlinearity. The corresponding master equation takes a similar form as Eq.~\eqref{eq:master}:
\begin{eqnarray}
\rho_\sigma(t + 1) = \sum_{\sigma'}p_{\sigma\sigma'}(t)\rho_{\sigma'}(t).
\label{mf_master_eq}
\end{eqnarray}

\begin{figure}[h]
    \centering
    \includegraphics[width=1.0\textwidth]{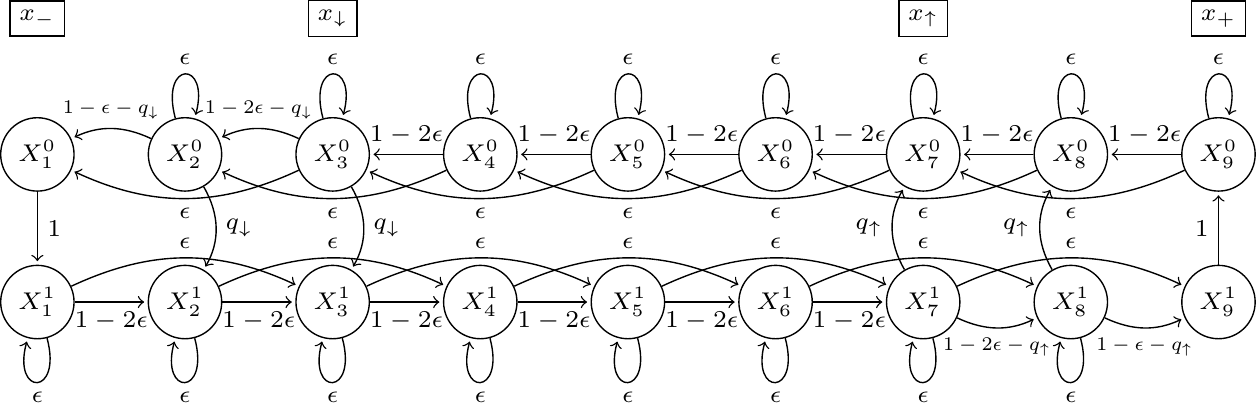}
    \caption{Full transition matrix $p_{\sigma\sigma'}(t)=p^{(0)}_{\sigma\sigma'}+ q_{\downarrow}(t)p^{\downarrow}_{\sigma\sigma'}+ q_{\uparrow}(t)p^{\uparrow}_{\sigma\sigma'}$, $0\leq q_{\downarrow}(t)\leq1-2\epsilon$, $0\leq q_{\uparrow}(t)\leq1-2\epsilon$.}
    \label{fig4}
\end{figure}

According to the graphical representation of the matrix $p_{\sigma\sigma'}(t)$ in Fig.~\ref{fig4}, each particular load may experience four types of transition while in the out-of-comfort zone: i/ it may remain in the same state with probability $\epsilon$; ii/ it may go one step deeper in the out-of-comfort zone with probability $1 - 2\epsilon - q_{\uparrow/\downarrow}(t)$ (where for ease of notation $\uparrow/\downarrow$ means either $\uparrow$ or $\downarrow$); iii/ it may go two steps deeper in the out-of-comfort zone with probability $\epsilon$; iv/ it may switch state from on (resp. off) to off (resp. on) with probability $q_{\downarrow}(t)$ (resp. $q_{\uparrow}(t)$). Each particular probability must be non-negative; so the Poisson rates must satisfy $q_{\uparrow/\downarrow}(t)\leq1-2\epsilon$. Consequently, $f(x)$ is restricted to the $[0; 1-2\epsilon]$ interval. Acknowledging that many choices are possible, we choose to work with the following form of the saturation function: 
\begin{equation}\label{f_function}
f(x) = \begin{cases}x,\ x<1-2\epsilon,\\1-2\epsilon,\ {\rm otherwise}\end{cases}
\end{equation} 
Note that the discrete schemes described by the transition matrices Eqs.~(\ref{eq:p-sum}) and (\ref{master_eq}) have a proper continuous limit, as the corresponding master equations transform into Fokker-Planck equations discussed in \cite{ Chertkov2017, Metivier2019, Metivier2020} in this limit. (See Appendix \ref{AppA} for details.)

To measure the system evolution toward steady state, we use two quantities: $H_1(t) = \|\rho^{\rm (st)}_\sigma - \rho_{\sigma}(t)\|_1$, which is the $L^1$ distance describing how the probability mass function $\rho_\sigma(t)$ goes toward its steady state; and the $|N_{\uparrow}(t) - N^{(\rm st)}_{\uparrow}|$, which provides a measure of how the total energy consumption of the ensemble goes toward its steady-state value set by the aggregator. We show below that in the case of the mean-field control the rate of the two quantities relaxation to the steady state may be dramatically different. Specifically, $|N_{\uparrow}(t) - N^{(\rm st)}_{\uparrow}|$ may converge to $0$ much faster than $H_1(t)$.
\subsection{Linear analysis of the decay rate} 

Standard eigenvalue analysis of the linear master equation Eq.~\eqref{eq:master} (with the constant switching rates) shows that there is a unique maximal eigenvalue equal to unity if $\epsilon>0$, so that the steady state $\rho^{\rm (st)}_\sigma$ is unique. Then, if $N_\uparrow^{(\rm{st})}=\frac{1}{2}$ (this is proven in Appendix \ref{AppB}), the nonlinear master equation Eq.~\eqref{mf_master_eq} also has the same steady-state, which is unique in a small neighborhood. In our numerical simulations, we did not encounter other steady states. To see how fast the system approaches its steady state with the mean-field control, we proceed with the eigenvalue analysis of Eq.~\eqref{mf_master_eq}. 

Applying the decomposition, $\rho_\sigma(t) = \rho^{(\rm st)}_\sigma+\delta \rho_\sigma(t)$, and keeping only the linear term we arrive at
\begin{eqnarray}
&&\delta \rho_\sigma(t+1) = \sum_{\sigma'}S_{\sigma\sigma'}\delta \rho_{\sigma'}(t),\nonumber\\
&&S_{\sigma\sigma'}=p_{\sigma\sigma'}^{(0)}+r p^{\downarrow}_{\sigma\sigma'}+ r p^{\uparrow}_{\sigma\sigma'}+2\alpha rU_{\sigma'} \sum_{\sigma''}p^{\downarrow}_{\sigma\sigma''}\rho_{\sigma''}^{(\rm{st})}-2\alpha rU_{\sigma'} \sum_{\sigma''}p^{\uparrow}_{\sigma\sigma''}\rho_{\sigma''}^{(\rm{st})},
\label{linear_master_eq}
\end{eqnarray}
where we have used the equality $N_{\uparrow}^{(\rm{st})}=1/2$. The transition matrix $S$ defined in Eq.~\eqref{linear_master_eq} can be split in two parts: $p$, which is the transition matrix of the ensemble without mean-field control, Eq.~\eqref{eq:p-sum}, and the term $V$, which can be treated as a perturbation. The matrix elements $S_{\sigma\sigma'}$, $p_{\sigma\sigma'}$, and $V_{\sigma\sigma'}$ read:

\begin{eqnarray}
&&S_{\sigma\sigma'}= p_{\sigma\sigma'} + V_{\sigma\sigma'},\nonumber\\
&&p_{\sigma\sigma'}=p_{\sigma\sigma'}^{(0)}+r p^{\downarrow}_{\sigma\sigma'}+r p^{\uparrow}_{\sigma\sigma'},\nonumber\\
&&V_{\sigma\sigma'}=2\alpha rU_{\sigma'} \sum_{\sigma''}p^{\downarrow}_{\sigma\sigma''}\rho^{\rm (st)}_{\sigma''}-2\alpha rU_{\sigma'} \sum_{\sigma''}p^{\uparrow}_{\sigma\sigma''}\rho^{\rm (st)}_{\sigma''}.
\end{eqnarray}

\noindent where $r$ is constant. The spectral decomposition of the transition matrix $S$ yields:

\begin{eqnarray}
S_{\sigma\sigma'}=\sum_{i}\Lambda^{(i)} \psi^{(i)}_\sigma\phi^{(i)}_{\sigma'},
\end{eqnarray}

\noindent where $\left\{\Lambda^{(i)}\right\}$ is the set of eigenvalues, and  $\left\{\psi^{(i)}_\sigma\right\}$ and $\left\{\phi^{(i)}_{\sigma'}\right\}$ are respectively the right eigenvectors and the left eigenvectors sets such that $\sum_{\sigma}\phi^{(i)}_\sigma\psi^{(j)}_\sigma=\delta_{ij}$. The time evolution of the perturbation $\delta \rho_\sigma(t)$ is then given by:

\begin{equation}
\delta\rho_\sigma(t) = \sum_\sigma\sum_i \left(\Lambda^{(i)}\right)^t \psi^{(i)}_\sigma\phi^{(i)}_{\sigma'} \delta\rho_{\sigma'}(0). 
\end{equation}

Note that the matrix $S$ is a real matrix, so if $\Lambda^{(i)}$ is an eigenvalue of $S$ then $\Lambda^{(i)*}$ is also an eigenvalue i.e. all complex eigenvalues are paired. 

As shown in Appendix \ref{AppD}, there is at least one distinct eigenvalue $\Lambda^{(0)}=1$ whose corresponding eigenvector, $\psi_\sigma^{(0)}$, characterizes a mode that does not decay towards the steady state with time, but whose amplitude is always equal to $0$, in order to satisfy the normalization condition: $\sum_\sigma\rho_\sigma=1$. The other modes associated with the right eigenvectors $\psi_\sigma^{(i)}$ ($i\neq 0$) decay as $\left(\Lambda^{(i)}\right)^t$. It is convenient to introduce the relaxation constant of the mode as the complex number $\lambda_i = -\log(|\Lambda^{(i)}|) + i\arg\left(\Lambda^{(i)}\right)$. With these notations we rewrite the dynamics of the perturbation as:

\begin{equation}
\delta\rho_\sigma(t) = \sum_{\sigma'}\sum_i\exp\left[-{\rm Re}(\lambda_i)t + i{\rm Im}(\lambda_i)t\right]~\psi^{(i)}_\sigma\phi^{(i)}_{\sigma'} \delta\rho_{\sigma'}(0).
\end{equation}

\noindent The relaxation constants introduced here are the space-time-quantized analogues to the  Fokker-Planck operator's eigenvalues of the continuous models \cite{Chertkov2017, Metivier2019, Metivier2020}.

From now on we focus only on the exponential rates with the smallest real part, dominating the long-time relaxation. In \cite{Metivier2019}, authors found two classes of modes when analysing the continuous version of the model with mean field control: one that does not contribute at all to the total energy $N_\uparrow$ and one that does. Following this idea we sort our modes $\psi_\sigma^{(i)}$ into two families $u_\sigma^{(i)}$ and $v_\sigma^{(i)}$:
\begin{itemize}
    \item[$-$] The family of vectors $u^{(i)}_\sigma$ for which we have $\sum_{\sigma'}V_{\sigma\sigma'}u^{(i)}_{\sigma'} = 0\Longleftrightarrow \sum_\sigma U_\sigma u^{(i)}_\sigma=0$, where $i$ denotes the label of eigenvectors in the set, which we call the \emph{ghost} family, $\{\rm GF\}$, as vectors of this set do not contribute, after the DR perturbation is applied, neither to the energy consumption $U$ (or equivalently, $N_\uparrow$) nor to respective relaxation. However, and unless degeneracy, these modes contribute to other observables, in particular the $H_1$ -- distance between  steady-state and the current time state. 
    \item[$-$] The family of vectors $v^{(i)}_\sigma$ for which we have $\sum_{\sigma'}V_{\sigma\sigma'}v^{(i)}_{\sigma'} \neq 0$, which we call the \emph{significant} family, $\{\rm SF\}$, as it influences the energy consumption and its relaxation, contributing as well to the relaxation of the entire ensemble. 
\end{itemize}

The whole family, $\{\rm WF\}$, of eigenvalues $\Lambda$, is the union of the ghost family and the significant family of the eigenvalues: $\{\rm WF\} = \{\rm SF\} \cup \{\rm GF\}$. A similar classification of eigenvalues of the Fokker-Planck operator eigenvalues was done for the continuous models \cite{Metivier2019, Metivier2020}. Using the definitions above we can now introduce the relaxation rate as $\underset{\lambda\in\{\rm SF\}/\lambda_0}{\min}{\{\rm Re}\lambda\}$ and reintroduce the relaxation rate for the entire ensemble as $\underset{\lambda\in\{\rm WF\}/\lambda_0}{\min}{\{\rm Re}\lambda\}$. The detailed comparison of these relaxation rates is performed in the next Section.

\section{Numerical Results and Discussion}
\label{sec:experiment}

\subsection{Relaxation Constants}

\begin{figure}[h]
    \centering
    \includegraphics[width=1.0\textwidth]{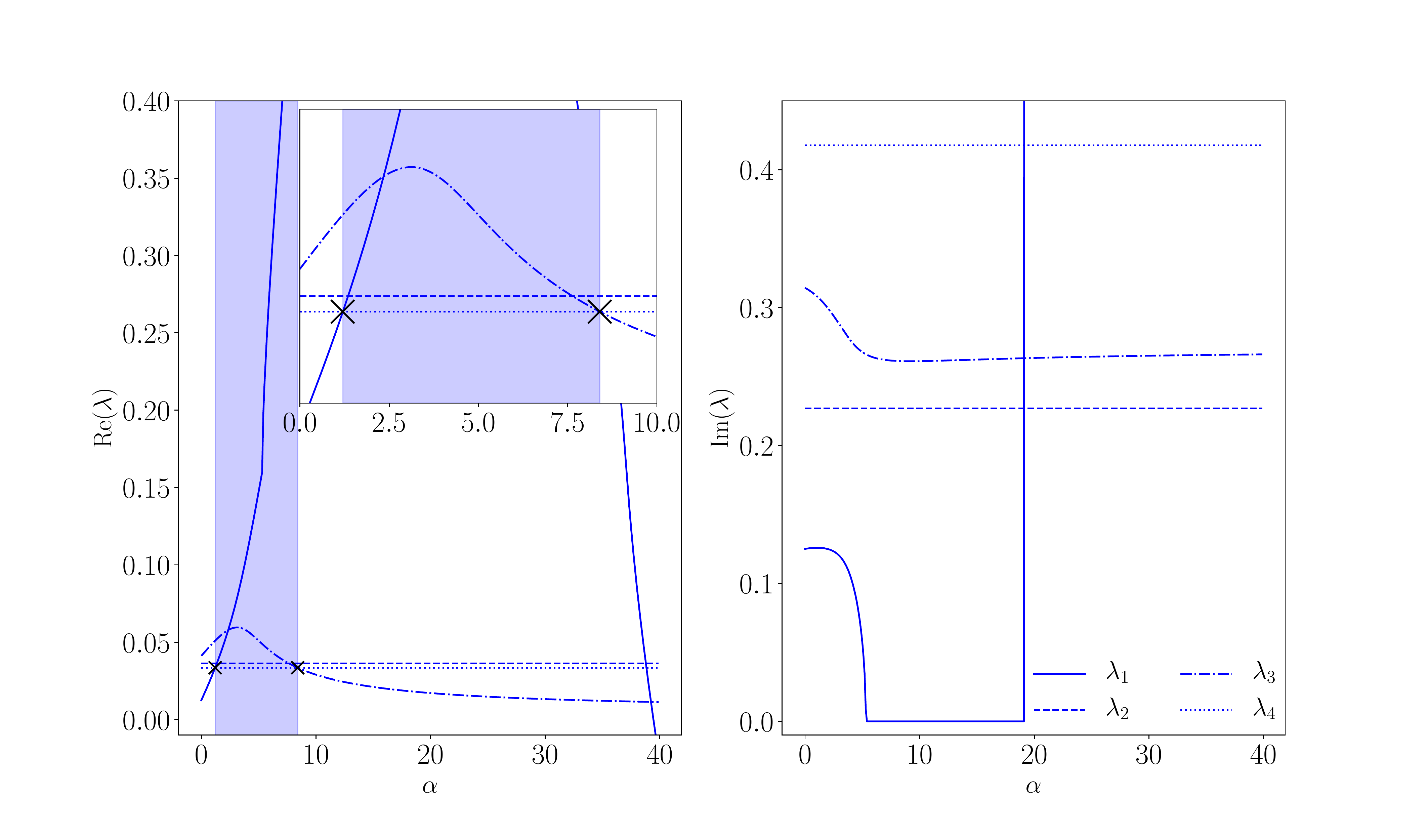}
    \caption{Real and imaginary parts, ${\rm Re}(\lambda_i)$ and ${\rm Im}(\lambda_i)$, of the first 4 relaxation constants as functions of the degree of nonlinearity $\alpha$ are shown for $r=0.1$. The number of states in the comfort and out-of-comfort zones are $n_{\rm in}=12$ and $n_{\rm out}=18$, respectively.
    }
    \label{fig5}
\end{figure}

We compute leading relaxation constants $\lambda_i$ numerically. The behavior of the first four constants, i.e. those with the smallest real parts, ${\rm Re}(\lambda_i)$, of the whole set excluding $\lambda_0 = 0$, is shown in Fig.~\ref{fig5} as functions of the strength of the mean-field signal, $\alpha$, for two different values of the Poisson rate, $r$. Eigenvalues associated with the ghost eigenvectors (even indexes) do not depend on $\alpha$ by definition, so the corresponding relaxation rates $\lambda_i$ are also $\alpha$-independent. Hence, only the eigenvalues of the significant family contribute relaxation of the consumption. The jump of ${\rm Im}(\lambda_i)$, seen on the right panel, is due to  the fact the the imaginary part is defined modulo $2\pi$. The blue stripe on the left panel marks the domain  where total consumption of the significant ensemble and the whole ensemble show different relaxation rates at $r=0.1$, i.e. ${\rm Re}(\lambda_2)<{\rm Re}(\lambda_1)$. The inset is a magnified view of the $[0;10] \times [0;0.75]$ domain in the $\{\alpha;{\rm Re}(\lambda)\}$ plane, with crosses mark intersection where the relaxation rates of the significant and ghost families start to deviate. Such domain does not exist at $r=0.3$.  

\subsection{Super-relaxation in space-time quantized model}
\begin{figure}[h]
    \centering
    \includegraphics[width=1.0\textwidth]{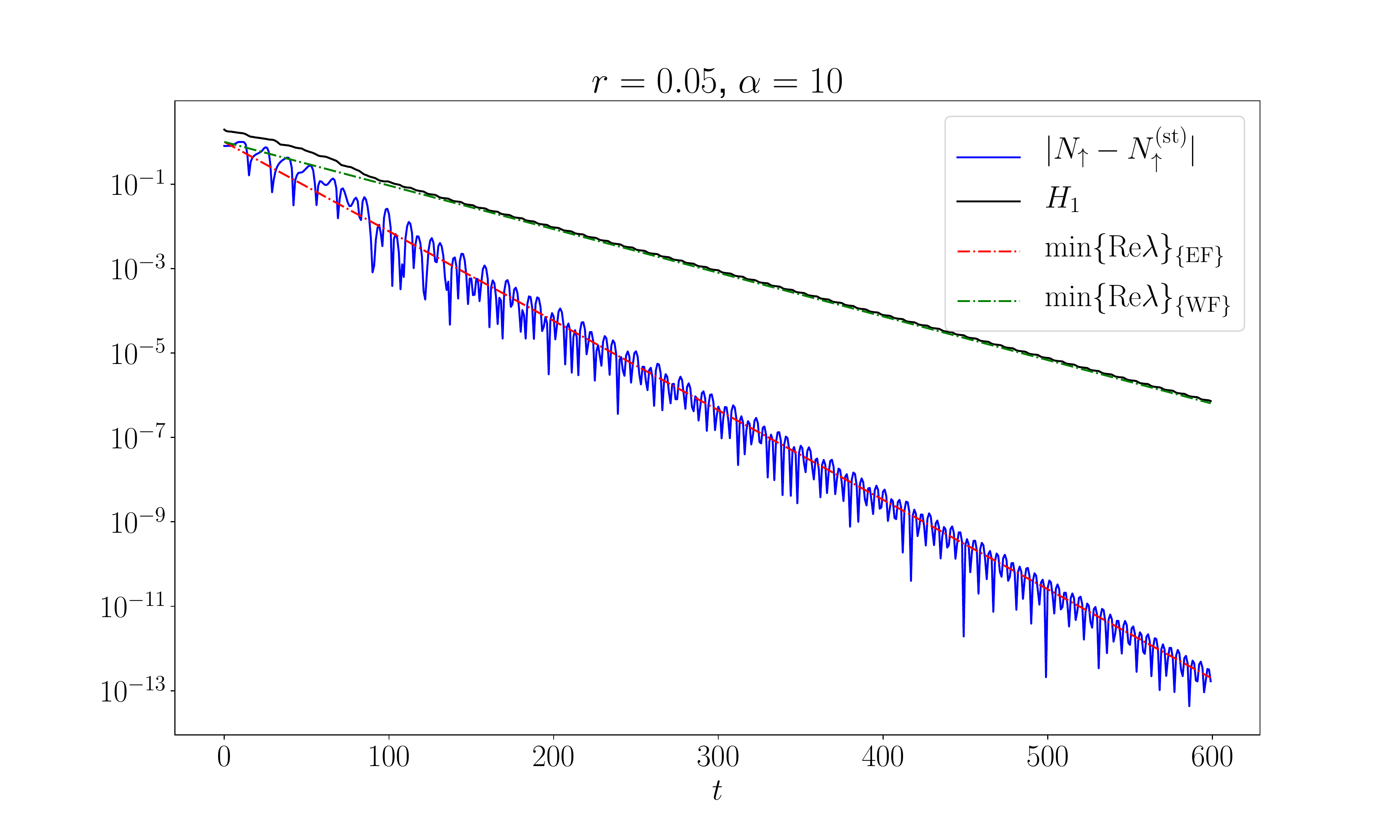}
    \caption{Illustration of the super-relaxation, with $\epsilon = 0.05$, $\alpha=10$ and $r=0.05$ by comparison of $|N_\uparrow (t)-N_\uparrow^{({\rm st})}|$ and $H_1 = \|\rho_{\sigma}(t) - \rho_{\sigma}^{(0)}\|_1$, reflecting how the whole ensemble relaxes to its steady state. The dashed-doted curves are the relaxation rates obtained by spectral decomposition. Both quantities decay as $e^{-\lambda t}$ with different $\lambda$.}
    \label{fig6}
\end{figure}

\begin{figure}
    \centering
    \includegraphics[width=1.0\textwidth]{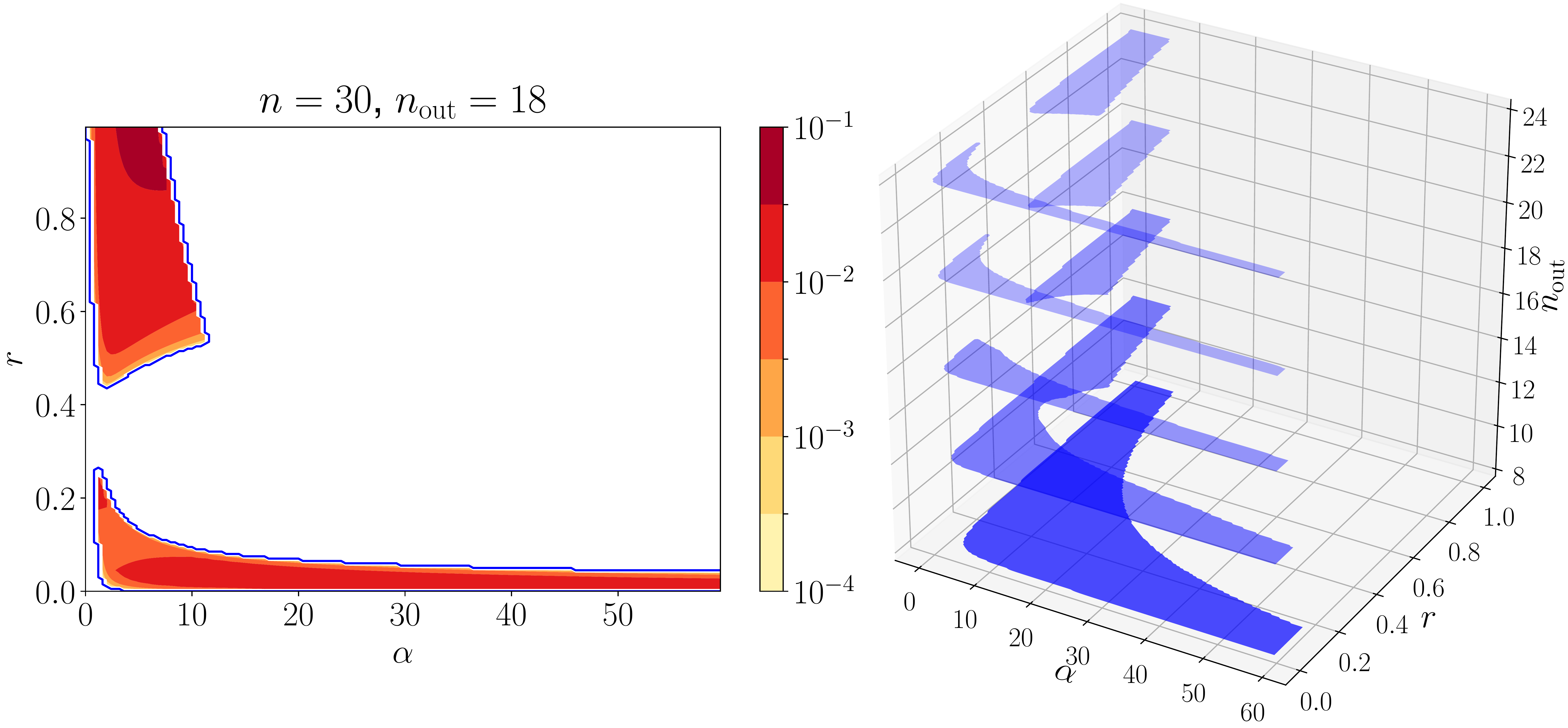}
    \caption{The left panel shows the gap ${\rm G}$ characterizing the super-relaxation regime, for different values of $r$ and $\alpha$, number of states in the comfort zone $n_{\rm in}$, number of states out of the comfort zone $n_{\rm out}$. The blue line marks the frontier between standard/super relaxation areas. The gap is zero across the whole white area. The right panel displays areas with $z$ coordinates that indicate the relaxation areas for various numbers of states in the out-of-comfort zone varying from $n_{\rm out} = 8$ to $n_{\rm out} = 24$ with the step $\Delta n_{\rm out} = 4$, while the number of states in the comfort zones is constant with $n_{\rm in} = 12$ in this example. Plots in both panels were produced with a diffusion coefficient $\epsilon = 0.05$.}
    \label{fig7}
\end{figure}

Super-relaxation is essentially the fast relaxation of the total consumption $N_\uparrow(t)$ while the $L^1$-distance $H_1(t)$ is much slower to reach steady state. Since only the significant family $\{\rm SF\}$ and its set of eigenvalues govern the relaxation of consumption, we may derive a criterion for the super-relaxation. In the general case, the relaxation rate $\lambda$ of the ensemble $\rho_\sigma$ takes the value $\min{{\rm Re}(\lambda)}$ as it yields the fastest characteristic decay time, while the relaxation rate for $N_\uparrow$ takes a different value: $\min_{\lambda\in \{{\rm EF}\}}{{\rm Re}(\lambda)}$. Mismatch between the two minima, $G$, is called the "gap":
\begin{equation}
{\rm G} =\min \{{\rm Re}(\lambda)\}_{\{\rm SF\}} - \min \{{\rm Re}(\lambda)\}_{\{\rm WF\}}.
\label{criterion}
\end{equation}
\noindent The gap determines the relaxation regime: standard or super-relaxation. If ${\rm G} = 0$ the system dynamics follows the standard regime; if ${\rm G} > 0$ the system undergoes super-relaxation. To better understand peculiarities of the super-relaxation regime in the space-time-quantized framework, we compute the ``phase diagram'' of the gap in the $\{\alpha;r\}$ plane, using Eq.~\eqref{criterion}. An illustrative example of the dynamics with super-relaxation is shown in Fig.~\ref{fig6}: $|N_\uparrow-N_\uparrow^{(\rm{st})}|$ goes to zero faster than $H_1$ does; we also see that $\min \{{\rm Re}(\lambda)\}_{\{\rm SF\}}$ and $\min \{{\rm Re}(\lambda)\}_{\{\rm WF\}}$  have different slopes ($\lambda_1$ and $\lambda_2$ respectively), which means that the gap G is nonzero. The two curves may cross each other thus closing their gap at some point in time. The particular point when the gap is zero (no super-relaxation) depends on both model parameters $r$ and $\alpha$. We analyze this further by calculating the phase diagram of G, showing two possible phases: standard relaxation and super-relaxation in Fig.~\ref{fig7}.

We denote $n$ the total number of states in up (down) position, and $n_{\rm out}$, the number of states in the out-of-comfort zone in up (down) position. For particular values of $n$ and $n_{\rm out}$, and a small diffusion coefficient $\epsilon$, the typical behavior of the gap as a function of both $r$ and $\alpha$ is shown in Fig.~\ref{fig7}. The super-relaxation domain in the $(\alpha,r)$ plane for different characteristics of interest in the out-of-comfort zones is also shown in Fig.~\ref{fig7}: for large values of $n_{\rm out}$, the super-relaxation $\{\alpha;r\}$-domain decreases significantly and tends asymptotically to a fixed shape as shown on the right panel of Fig.~\ref{fig7}. Convergence to the fixed shape is rather fast due to the fact that the probability mass is localized around the comfort zone, and that the far-lying out-of-comfort zone solutions do not influence dynamics of the model. These domains overlap: the deep orange domain is partly covered by the other domains, which are smaller in sizes. This result is consistent with the fact that the ensemble mixing, which favors fast relaxation, can hardly be achieved if the number of states in the out-of-comfort zone remains high. We have checked numerically that variation of the diffusion coefficient $\epsilon$, has a rather limited impact on the super-relaxation surface. We also verified that in the limit of infinite number of states in the out-of-comfort and comfort zones, value of the converges to the one correspondent to the continuous model, thus implying that the dynamical behavior described in Ref. \cite{Metivier2020} is recovered in this limit.

\subsection{Undamped oscillations in consumption} 

\begin{figure}[h]
    \centering
    \includegraphics[width=1.0\textwidth]{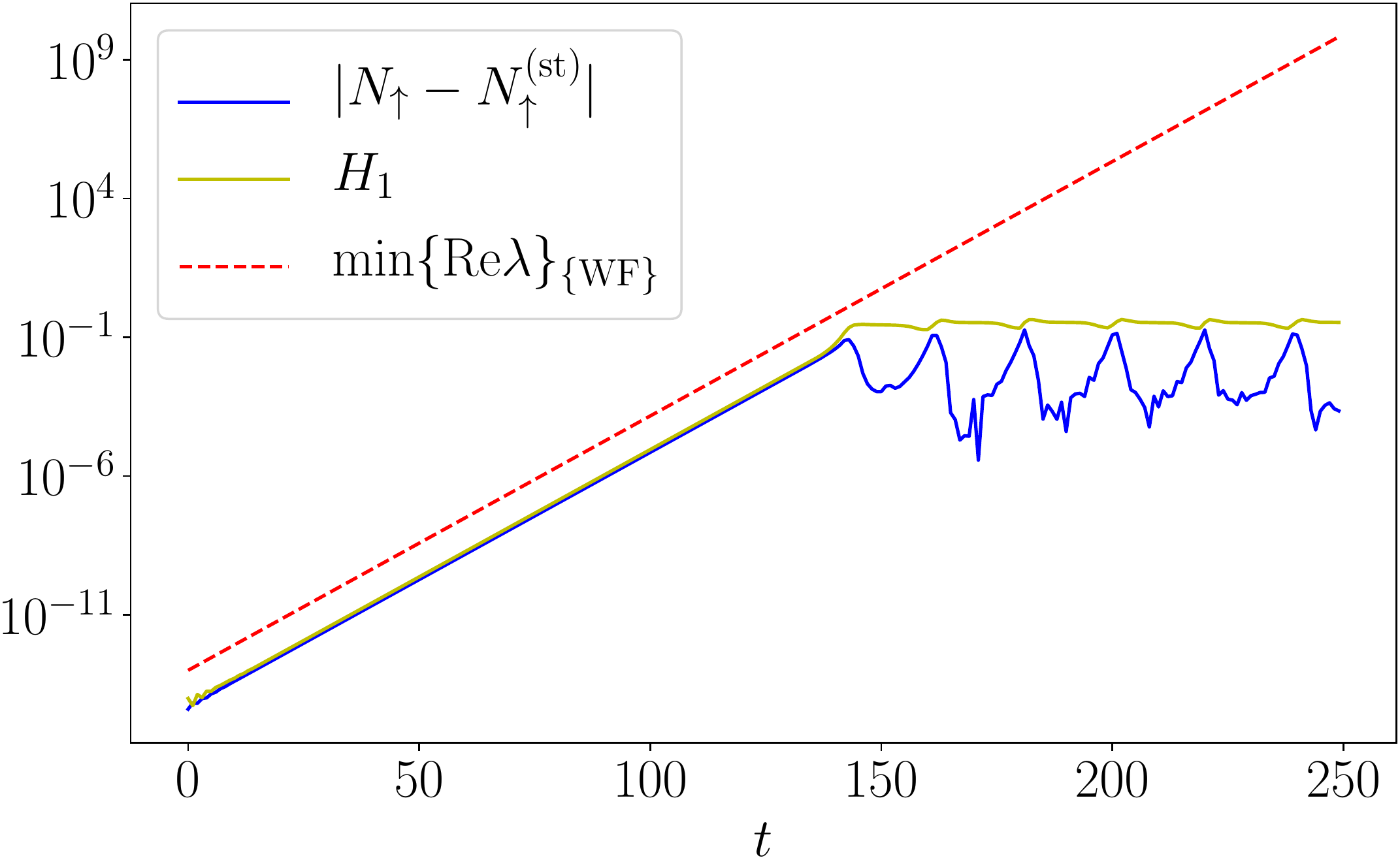}
    \caption{System dynamics with unstable steady state, as observed for $n = 30$, $n_{\rm out} = 18$, $r=0.2$ and $\alpha=25$. In this case $N_\uparrow$ and $H_1$ grow in time according to $e^{-\lambda t}$ as $\lambda$ is negative. Note that at some point, the exponential growth stops and the dynamics stabilizes; from this point on the linear analysis (red-dashed curve) no longer applies.}
    \label{fig8}
\end{figure}

As seen in Fig.~(\ref{fig8}), reporting experimental observations, at some values of $r$ and $\alpha$, and depending on how time discretization is implemented, undamped oscillations in consumption are observed. The oscillations are also preceded by the period of growth. 

This is clearly an undesirable phenomenon which needs to be explained. In the following we are discussing results of comparison of the experiments with nonlinear system juxtaposed against the linear stability analysis. The comparison shows that there exist a range of parameters where the linear analysis shows an instability fully consistent with the amplitude growth observed in the experiment. The oscillations are seen in the regime which is beyond the linear stability analysis.  Some details and discussions of the phenomenon are discussed in the following.

To understand better dynamics of the energy consumption after a DR perturbation we ought to monitor for super-relaxation but also for instability, checking the dominant relaxation rate, $\lambda_i$, i.e. one with the smallest real part. Contrary to the continuous model where only one crossing (correspondent to equal real parts of $\lambda_1$ and $\lambda_2$) is observed as we change $\alpha$, in the discrete case multiple events of level crossings are possible, e.g. as illustrated in Fig.~\ref{fig5}. This means that the entire spectrum of the  relaxation constants, $\{\lambda_i\}$, need to be considered to resolve which mode dominates the relaxation. 

\begin{figure}
    \centering
    \includegraphics[width=1.0\textwidth]{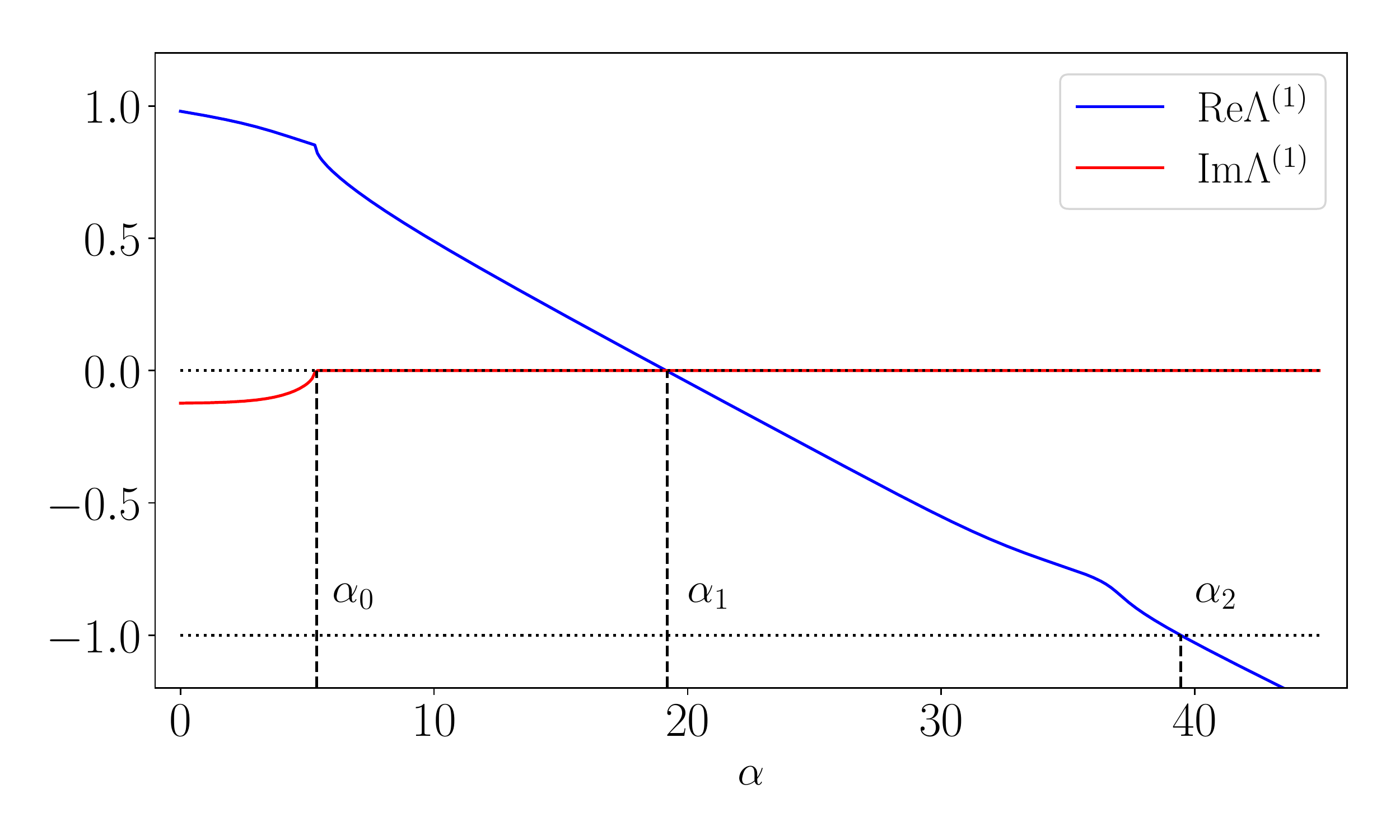}
    \caption{Real and imaginary parts of the eigenvalue $\Lambda^{(1)}$ as functions of the degree of nonlinearity $\alpha$. Here, $\Lambda^{(1)}$ becomes a real number from $\alpha_0 = 5.38$. The particular values at which $\Lambda^{(1)} = 0$ and $\Lambda^{(1)}$ becomes smaller than $-1$ are $\alpha_1 = 19.21$ and $\alpha_2 = 39.45$ respectively.}
    \label{fig9}
\end{figure}

Consider the case depicted in Fig.~\ref{fig9} and follow, as $\alpha$ varies, the peculiar behavior of the eigenvalue $\Lambda^{(1)}$, which is related to the relaxation constant, $\lambda_1$, according to $\lambda_1 = -\log(|\Lambda^{(1)}|) + i\arg (\Lambda^{(1)})$. Since at each time step the amplitude of the corresponding mode is multiplied by $\Lambda^{(1)}$, the mode decays in time if $|\Lambda^{(1)}|<1$, and it grows if $|\Lambda^{(1)}|\geq 1$. At sufficiently small $\alpha$ and before $\alpha$ reaches the value $\alpha_1$, $\alpha\leq \alpha_1$, the mode decays. (Notice that there is also another special value, $\alpha_0$, where $0<\alpha_0<\alpha_1$, such that $\mbox{Im}(\Lambda^{(1)})$ is finite at $\alpha<\alpha_0$ and it is zero at $\alpha\geq 0$. Crossing $\alpha_0$ does not have implications on how the mode decays.)
The aforementioned instability occurs when $\alpha$ becomes larger than $\alpha_2$ at which point $\Lambda^{(1)}=-1$.

\begin{figure}[h]
    \centering
    \includegraphics[width=1.0\textwidth]{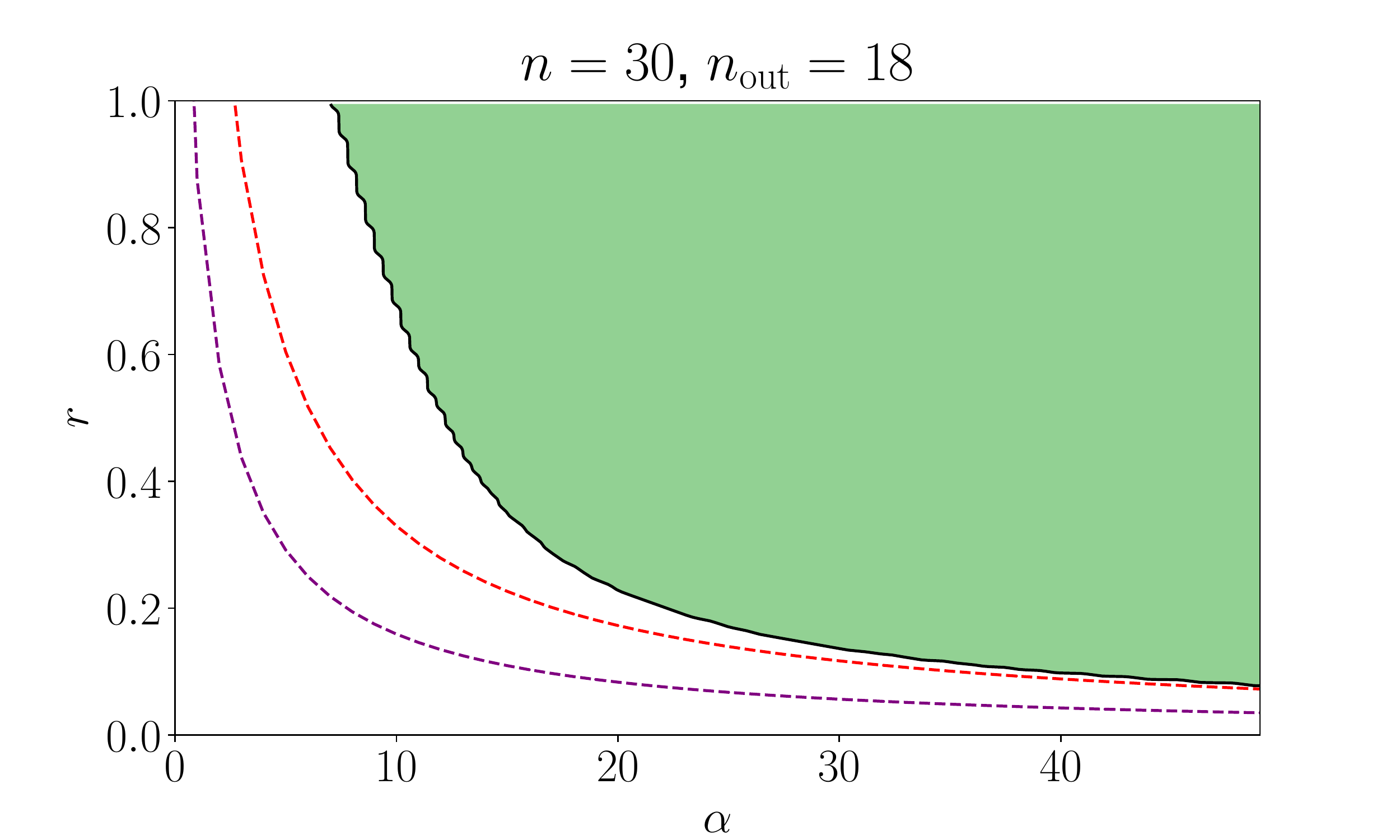}
    \caption{Phase diagram showing instability observed for $n = 30$ and $n_{\rm out} = 18$. The white zone, where the real part of all relaxation constants is positive, is stable. The green zone is unstable with at least one relaxation constant with negative real part. The red dashed curve shows the analytical estimation of $\alpha_2$ for different values of $r$, Eq.~\eqref{frontier}, approximately marking the frontier separating the two regimes. The dashed purple curve is the analytical estimation of the value of $\alpha_1$ for different $r$ that marks the frontier between dynamical regimes.}
    \label{fig10}
\end{figure}

It is useful to have a simple, albeit not absolutely precise, criterion which allows to avoid undesirable  instability following by oscillations. We suggest a criterion based on estimations of $\alpha_1$ and $\alpha_2$. As shown in Appendix \ref{AppC}, considering a simplified version of the dynamical equation, Eq.~\eqref{dynamicEq}, yields $\alpha_1\approx \frac{n-2}{r(n_{\rm out}-2)}-1$ and $\alpha_2\approx \frac{2(n-1)}{r(n_{\rm out}-2)}-1$. The estimation results in the following estimation of the frontier separating stable and unstable regimes (see Appendix \ref{AppC} for details)
\begin{equation}\label{frontier}
    \frac{n-2}{n} - 2\frac{n_{\rm out}/2-1}{n}(1+\alpha)r + 1 = 0
\end{equation}
Summary of the behavior, illustrating frontier (red dashed curve) where the instability occurs, is also shown in Fig.~\ref{fig10} for an exemplary values of the diffusion coefficient and the size of the out-of-comfort zone. We observe that only eigenvalues from the main family can lead to instability since the ghost family is not affected by the mean-field feedback. Even though the (red dashed) boundary correspondent to the criterion (\ref{frontier}) is not precise it nevertheless gives a conservative guidance on the range of parameters where the instability can be safely avoided. We conclude emphasizing that the instability is an unfortunate artifact of the discrete regime and it does not occur in the continuous regime discussed in \cite{Metivier2020}.

\subsection{Remarks on the linearity of $x(t)$}
To demonstrate super-relaxation in a model reflecting the discrete nature of an actual TCL ensemble, the relevant parameters to consider are the degree of nonlinearity $\alpha$, the Poisson rate $r$, and the number of load states within and outside the comfort zone, $n_{\rm in}$ and $n_{\rm out}$ respectively. Our space-time quantized model and hence our numerical results depend on the strong assumption of the linearity of $x(t)$, the impact of which can be quantified with a concrete example. Note first that since we work in the limit where $x(t)$ is linear, the dynamics is invariant under a shift of all temperature parameters by some constant $\delta x$, i.e. the transformation $x_{\uparrow(\downarrow)} \rightarrow x_{\uparrow(\downarrow)} + \delta x$ does not change the dynamics. There is thus no need for an explicit knowledge of the temperature threshold values, especially as we aim for generality of our objective: observation of super-relaxation in discrete models pertinent to real-life cases. 

Now, let us consider the following illustrative case of a hot summer period when air conditioners are operating at high power. The outside temperature is $x_+=32^{\circ}$C, the minimal possible indoor temperature with a conditioner constantly being in the on position is $x_-=12^{\circ}$C, and the indoor comfortable temperature zone goes from $x_{\downarrow} = 21^{\circ}$C to $x_{\uparrow} = 23^{\circ}$C, with the mean being $x_{\rm c} = 22^{\circ}$C. The temperature dynamics, when the conditioner is in its off state, satisfies $\frac{dx(t)}{dt}=\zeta (x_{+} - x(t))$, where $\zeta$ is some variation rate. One can decompose $x(t)$ as follows $x(t) = x_{\rm c} + \Delta x(t)$, where $\Delta x(t)$ is a deviation of $x(t)$ about a mean comfortable temperature $x_{\rm c}$. Then, we obtain $\frac{d\Delta x(t)}{dt} = \zeta(x_{+} - x_c - \Delta x(t))$. The solution of this equation reads
$\Delta x(t) - \Delta x(0) = \zeta(x_+ - x_c) t - \zeta \int_0^t \Delta x(\tau)d\tau$. 
The second term on the r.h.s. is the nonlinear term in $t$ that is neglected when assuming linear evolution in Eq.~\eqref{eq1.}. 
$\Delta x(t)$ is at most of order $\frac{x_\uparrow - x_\downarrow}{2}$ (for large switching rates which is our case). Comparing the linear and nonlinear terms, we find
$\dfrac{\int_0^t \Delta x(\tau)d\tau }{(x_+ - x_c) t}\lesssim \dfrac{x_\uparrow - x_\downarrow}{2(x_+ - x_c)} = 1/10
$
for the realistic parameters validating our linear dynamics assumption. Similarly for the on state we get 
$
    \dfrac{\int_0^t \Delta x(\tau)d\tau }{(x_c- x_-) t}\lesssim \dfrac{x_\uparrow - x_\downarrow}{2(x_c - x_-)} = 1/10 \ll 1.
$
Note that in this illustrative example both cooling and heating rates give 1/10, which shows at the same time that the symmetric rate assumption is also valid.

\subsection{Influence of the Poisson rate on the consumers' comfort}
From the consumers' viewpoint, remaining in the comfort zone is the most important aspect of the problem. Generally speaking, comfort can been seen as based on a subjective perception of indoor microclimate, but one may still use objective parameters and values such as indoor temperature \cite{ashrae,Fanger1970}, humidity, CO$_2$ concentration \cite{Satish2012,Allen2016}, etc. to define bounds for comfort in the frame of a model. In our framework, the level of comfort can be defined as $C(t) = \sum_\sigma \rho_\sigma(t)C_\sigma$, where $C_\sigma$ is the weight of a particular state from the consumers' comfort point of view: $C_\sigma = 1$ in the comfort zone, and 0 outside. In the regime when the stationary state of the ensemble is slightly perturbed because of DR, the dynamics of $C(t)$ reads
\begin{equation}
    C(t) = \sum_i\left(\Lambda^{(i)}\right)^t \sum_\sigma \psi^{(i)}_\sigma C_\sigma \sum_{\sigma} \phi^{(i)}_\sigma \delta \rho_{\sigma}(0) + C^{(\rm st)},
\end{equation}
where $C^{(\rm st)} = \sum_{\sigma}\rho_\sigma^{(\rm st)} C_\sigma$.

\begin{figure}[h]
    \centering
    \includegraphics[width=1.0\textwidth]{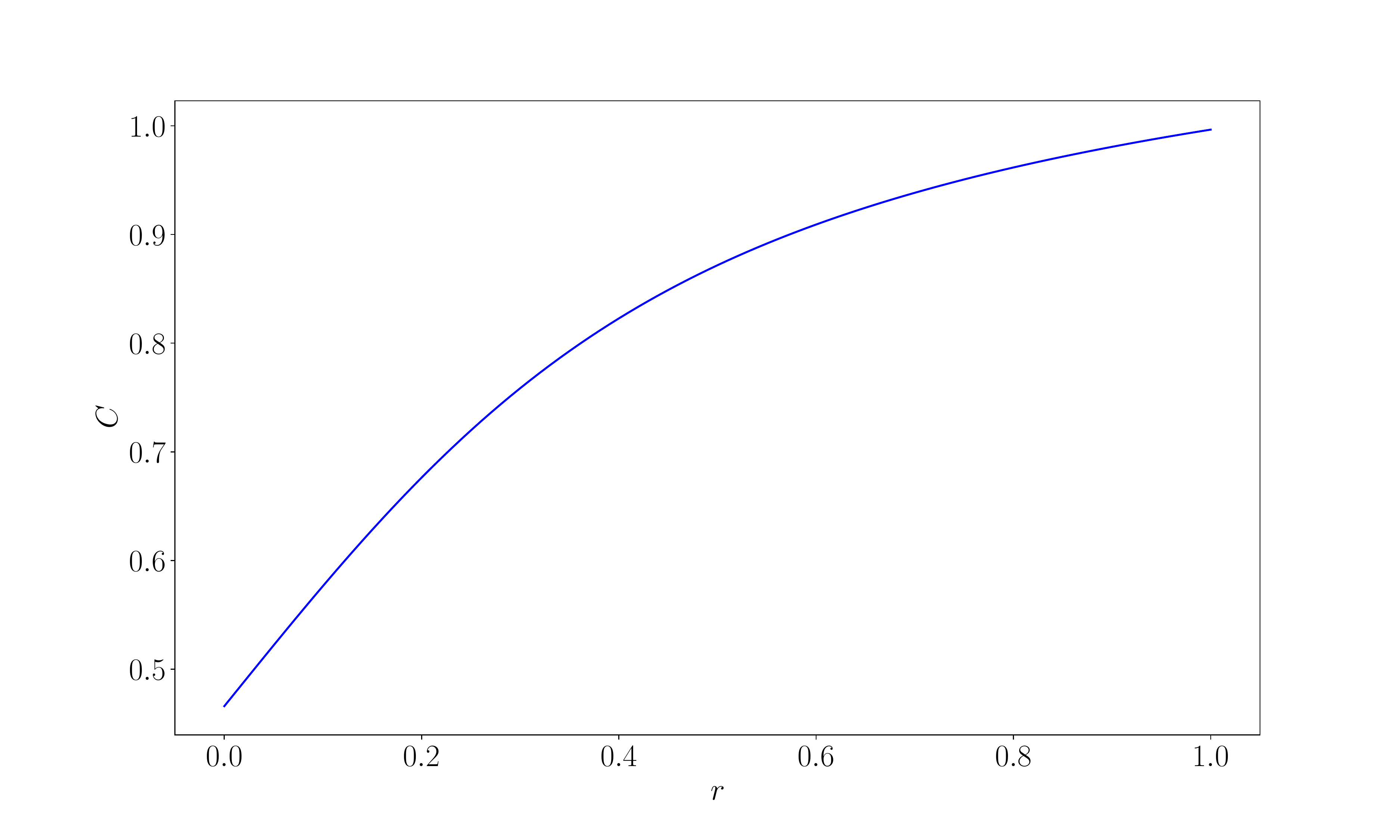}
    \caption{Dependence of the comfort level $C$ on the Poisson rate $r$.}
    \label{fig11}
\end{figure}

In our work we are interested in air conditioners as part of a large ensemble, and the key parameter to consider is temperature. The air conditioners are in our model two-state (on/off) devices subjected to randomized switching characterized by the Poisson rate $r$. From the power systems viewpoint, the main goal is to recover a sufficient level of ensemble mixing to avoid oscillations after DR perturbation. The smaller $r$ is the faster the mixing is, but this obviously implies an increased Poisson time ($1/r$) and the resulting delayed switching on the level of consumers comfort as a deviation of temperature from thresholds as illustrated in Fig.~\ref{fig11}. This particular point was addressed and discussed in \cite{Chertkov2017}.

\begin{figure}[h]
    \centering
    \includegraphics[width=1.0\textwidth]{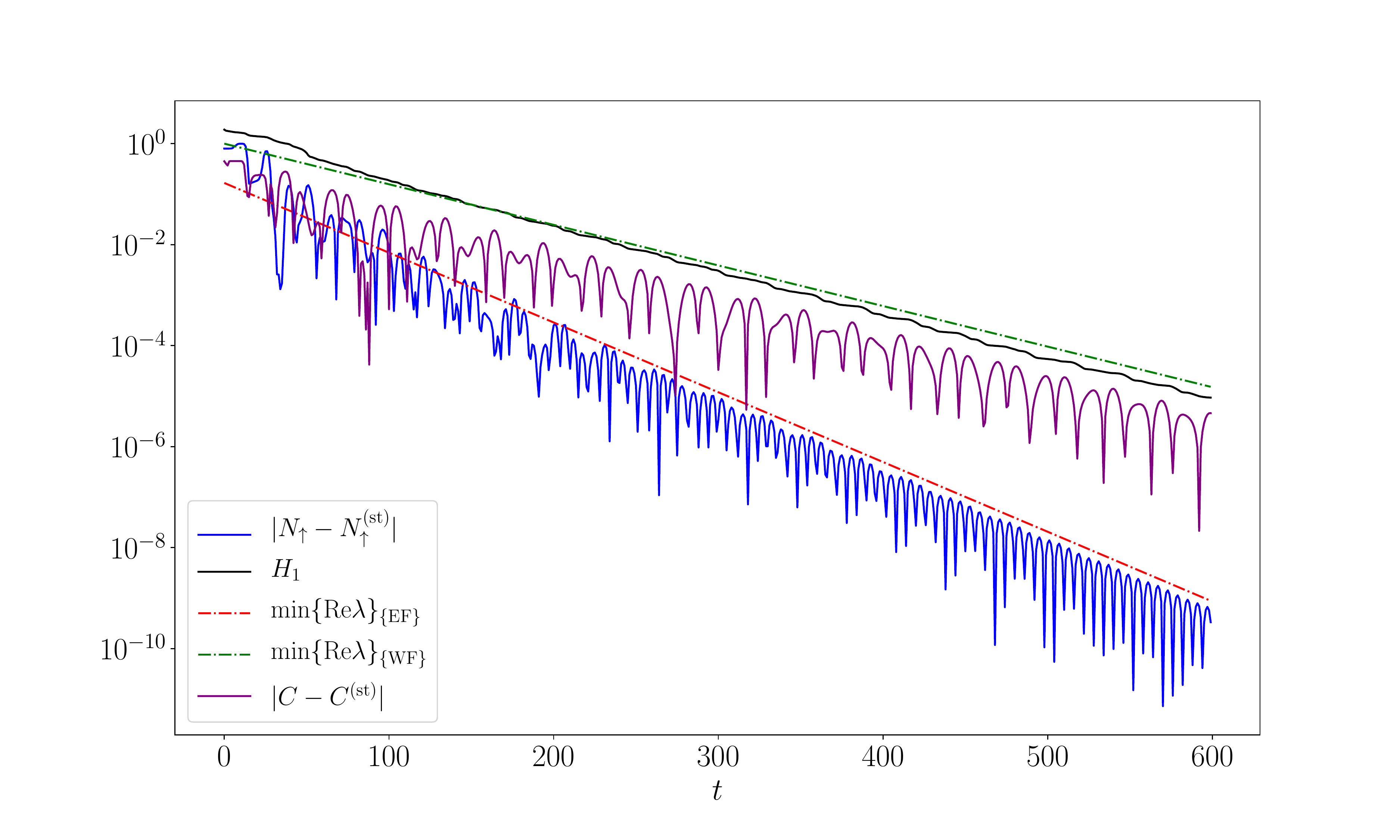}
    \caption{Temporal dynamics of different observable of the ensemble. $\alpha = 10$, $r=0.05$.}
    \label{fig12}
\end{figure}

We may also comment on the impact of non-linearity, if it were to be introduced, on the consumers' comfort level. One can see that in the general case, the consumers' comfort relaxes to its stationary value obeying the same relaxation constants as other observables of the ensemble. Indeed, as shown in Fig.~\ref{fig12}, the consumers comfort relaxes with the same rate as $H_1$, the $L^1$ distance describing how the probability mass function $\rho_{\sigma}(t)$ goes toward its steady state. Hence, introduction of non-linearity would not have any impact on the level of consumers comfort in the steady-state, because it does not have any impact on the steady-state.

\section{Conclusions and path forward}
\label{sec:conclusions} 

We start the concluding Section of the manuscript with a brief summary of the results reported:

\begin{itemize}
    \item Effect of the super-relaxation, previously observed in the continuous time model, extends to more realistic discrete time models where it becomes a useful practical tool for demand response. 
    
    \item We show that the super-relaxation is stable with respect to variations, fluctuations and uncertainty in the operationally sensible range of the model parameters.
    
    \item We also observe that dynamics of the TCL ensemble is sensitive to some details of the discretization scheme. In particular, for values of the ensemble parameters correspondent to large accumulations of nonlinear effects (including feedback) over a time step the system becomes linearly unstable then resulting in parasitic oscillations. We analyze the instability and provide a simple to implement criteria which allows to avoid the undesirable regime. 
\end{itemize}

Discussing the last point in some extra details, it is important to emphasize that emergence of the parasitic instability is a special feature of the discrete time model not observed in the continuous time model. We observed  that the super-relaxation in space-time-quantized models entails a more complicated spectral structure than that obtained with the continuous model \cite{Metivier2020}. We saw that undamped oscillations may arise if the discretization scheme is not calibrated proper. To uncover this effect we perform linear stability analysis and establish criteria for instability, then suggesting criteria on how to avoid it. This instability analysis allows us to claim that the manuscript contributes to the growing body of work towards establishing regimes for safe operations of the TCL ensembles, i.e. seeking for operations which allow to mitigate various parasitic effects. It is important to emphasize, however, that the oscillations reported in this manuscript  are not related to (but rather imposed on the top of) other oscillations already discussed in the literature and associated with irregular patterns of consumption and synchronization following demand response signals \cite{Sinitsyn2012,Mehta2014}. We conclude that aggregators and other participants of the energy markets should be aware of this newly reported  discretization-caused instability as it may be dangerously enhanced, if not mitigated proper, in the case of increasing level of fluctuations caused,  for example, by increase of renewable penetration.

Let us now turn to a brief discussion of the path forward. Even though the paper constitutes a significant step towards realistic operation of TCL ensembles, more work is needed to adapt our results to practical setting of the demand response implementations. We envision relaxing various assumptions made in this study to simplify the analysis, such as accounting for asymmetry in heating and cooling, accounting for variations of parameters on the level of individual devices, etc. More detailed physical modeling at the device scale may and should  include in the future modeling and monitoring  of the air-quality, i.e. CO$_2$ concentration, particulate matter concentration, aerosols, humidity, e.g. as discussed in \cite{Ryzhov2019}. 
 
Finally, we would like to emphasize that demand response is a general energy management tool, which is not restricted to power systems, and is in fact of an even greater utility for integrated energy systems \cite{Liu2020,Hassan2020}. The mean-field approach may also be extended to other infrastructure systems such as battery-, water-, waste-, and oil-product systems dependent on flexible consumers engaged in communications-light demand-response services. 

\section*{Acknowledgments} 
The work at LANL was carried out under the auspices of the National Nuclear Security Administration of the U.S. Department of Energy under Contract No. DE-AC52-06NA25396, and it was partially supported by DOE/OE/GMLC and LANL/LDRD/CNLS projects. The work at Skoltech was supported by the Skoltech NGP Program (Skoltech-MIT joint project).

\appendix

\section{Continuous limit of discrete master equation}
\label{AppA}

Here we discuss the master equation describing an ensemble of loads in continuous space-time limit. We parameterize the state of a device by the tuple, $\{x, \gamma\}$, where $\gamma = \uparrow, \downarrow$ marks the state of a load and $x$ marks the indoor temperature. Let us denote the temperature difference between neighbouring nodes, $\Delta x$, the number of nodes out of the comfort zone $n_{\rm out}$, the number of nodes in the comfort zone $n_{\rm in}$, the total number of nodes $n=n_{\rm out}+n_{\rm in}$, and the discrete time step $\Delta t$. We assume that $n_{\rm out}\gg n_{\rm in}$, i.e. $n_{\rm out}\rightarrow\infty$. Then, the discrete in space and time master equation \eqref{eq:master} takes the following form:

\begin{eqnarray}\label{eq:master-cont-app}
\begin{cases}
\rho_{\uparrow}(x, t + \Delta t) = (1 - 2\epsilon)\rho_{\uparrow}(x + \Delta x, t) + \epsilon \rho_{\uparrow}(x + 2\Delta x, t) + \epsilon\rho_{\uparrow}(x, t), \ x_{\downarrow}\leq x <x_{\uparrow},\\
\rho_{\downarrow}(x, t + \Delta t) = (1 - 2\epsilon)\rho_{\downarrow}(x - \Delta x, t) + \epsilon \rho_{\downarrow}(x - 2\Delta x, t) + \epsilon\rho_{\downarrow}(x, t), \ x_{\downarrow} < x \leq x_{\uparrow},\\
\rho_{\uparrow}(x, t + \Delta t) = (1 - 2\epsilon - q_{\downarrow}(t))\rho_{\uparrow}(x + \Delta x, t) + \epsilon \rho_{\uparrow}(x + 2\Delta x, t) + \epsilon\rho_{\uparrow}(x, t), \ x < x_{\downarrow},\\
\rho_{\downarrow}(x, t + \Delta t) = (1 - 2\epsilon - q_{\uparrow}(t))\rho_{\downarrow}(x - \Delta x, t) + \epsilon \rho_{\downarrow}(x - 2\Delta x, t) + \epsilon\rho_{\downarrow}(x, t), \ x_{\uparrow} < x,\\
\rho_{\uparrow}(x, t + \Delta t) = (1 - 2\epsilon)\rho_{\uparrow}(x + \Delta x, t) + q_{\uparrow}(t)\rho_{\downarrow}(x, t) + \epsilon \rho_{\uparrow}(x + 2\Delta x, t) + \epsilon\rho_{\uparrow}(x, t), \ x_{\uparrow}\leq x,\\
\rho_{\downarrow}(x, t + \Delta t) = (1 - 2\epsilon)\rho_{\downarrow}(x - \Delta x, t) + q_{\downarrow}(t)\rho_{\uparrow}(x, t) + \epsilon \rho_{\downarrow}(x - 2\Delta x, t) + \epsilon\rho_{\downarrow}(x, t), \ x\leq x_{\downarrow}.
\end{cases}
\end{eqnarray}

One may study this system of equations with discrete derivatives denoted as $D_x^{(n)}$, where $n$ is the order of a derivative and $x$ is the target variable. For example $D_x^{(1)}(F(x, y)) = \frac{F(x + \Delta x, y) - F(x, y)}{\Delta x}$ or $D^{(2)}_x(F(x, y)) = \frac{F(x - \Delta x, y) + F(x + \Delta x, y) - 2F(x, y)}{\Delta x^2}$. In these new notations the system of Eqs.~(\ref{eq:master-cont-app}) becomes 

\begin{eqnarray}
\begin{cases}
D_t^{(1)}(\rho_\uparrow(x, t)) = \frac{\Delta x}{\Delta t}D_x^{(1)}(\rho_\uparrow(x, t)) + \frac{\Delta x^2\epsilon}{\Delta t}D_x^{(2)}(\rho_\uparrow(x + \Delta x, t)), \ x_\downarrow \leq x < x_\uparrow,\\
D_t^{(1)}(\rho_\downarrow(x, t)) = -\frac{\Delta x}{\Delta t}D_x^{(1)}(\rho_\downarrow(x - \Delta x, t)) + \frac{\Delta x^2\epsilon}{\Delta t}D_x^{(2)}(\rho_\downarrow(x - \Delta x, t)), \ x_\downarrow < x \leq x_\uparrow,\\
D_t^{(1)}(\rho_\uparrow(x, t)) = \frac{\Delta x}{\Delta t}D_x^{(1)}(\rho_\uparrow(x, t)) + \frac{\Delta x^2\epsilon}{\Delta t}D_x^{(2)}(\rho_\uparrow(x + \Delta x, t)) - \frac{q_\downarrow(t)}{\Delta t}\rho_\uparrow(x + \Delta x, t), \ x < x_\downarrow,\\
D_t^{(1)}(\rho_\downarrow(x, t)) = -\frac{\Delta x}{\Delta t}D_x^{(1)}(\rho_\downarrow(x - \Delta x, t)) + \frac{\Delta x^2\epsilon}{\Delta t}D_x^{(2)}(\rho_\downarrow(x - \Delta x, t)) - \frac{q_\uparrow(t)}{\Delta t}\rho_\downarrow(x - \Delta x, t), \ x_\uparrow < x,\\
D_t^{(1)}(\rho_\uparrow(x, t)) = \frac{\Delta x}{\Delta t}D_x^{(1)}(\rho_\uparrow(x, t)) + \frac{\Delta x^2\epsilon}{\Delta t}D_x^{(2)}(\rho_\uparrow(x + \Delta x, t)) + \frac{q_\uparrow(t)}{\Delta t}\rho_\downarrow(x, t), \ x_\uparrow \leq x,\\
D_t^{(1)}(\rho_\downarrow(x, t)) = -\frac{\Delta x}{\Delta t}D_x^{(1)}(\rho_\downarrow(x - \Delta x, t)) + \frac{\Delta x^2\epsilon}{\Delta t}D_x^{(2)}(\rho_\downarrow(x - \Delta x, t)) + \frac{q_\downarrow(t)}{\Delta t}\rho_\uparrow(x, t), \ x \leq x_\downarrow.
\end{cases}
\end{eqnarray}

Let us now consider the limit
\begin{eqnarray}
&&n_{\rm in} \rightarrow \infty,\nonumber\\
&&\Delta x \rightarrow 0,\nonumber\\
&&\epsilon = {\rm const}\in [0, 0.5]\nonumber\\
&&\Delta t \rightarrow 0, \nonumber\\
&&r \rightarrow 0 \nonumber\\
&&n_{\rm in}\Delta x = {\rm const} = L,\nonumber\\
&&\frac{\Delta x}{\Delta t} = {\rm const} = v, \nonumber\\
&&\frac{r}{\Delta t} = {\rm const} = r_{\rm c}\ (\text{the subscript c refers to the continuous case.})
\end{eqnarray}
Notice that diffusion related term, $O(\sqrt{\Delta t})$, vanishes in the limit. Also,  there is no longer a need, in this limit, for the function $f$, whose role in the quantized model was to ensure non-negativity of the transition matrix elements. This discrete in space and time master equation turns into the following system of continuous Fokker-Planck equations, Eq.~\eqref{eq3.}:
\begin{eqnarray}
\begin{cases}
\frac{\partial\rho_\uparrow(x,t)}{\partial t} = v\frac{\partial \rho_\uparrow(x, t)}{\partial x}, \ x_\downarrow \leq x < x_\uparrow,\\
\frac{\partial\rho_\downarrow(x,t)}{\partial t} =  -  v\frac{\partial \rho_\downarrow(x, t)}{\partial x}, \ x_\downarrow < x \leq x_\uparrow,\\
\frac{\partial\rho_\uparrow(x,t)}{\partial t} = v\frac{\partial \rho_\uparrow(x, t)}{\partial x} - 2r_{\rm c}(N_\uparrow(t))^\alpha\rho_\uparrow(x, t), \ x < x_\downarrow,\\
\frac{\partial\rho_\downarrow(x,t)}{\partial t} = - v\frac{\partial \rho_\downarrow(x, t)}{\partial x}- 2r_{\rm c}(1 - N_\uparrow(t))^\alpha\rho_\downarrow(x, t), \ x_\uparrow < x,\\
\frac{\partial\rho_\uparrow(x,t)}{\partial t} = v\frac{\partial \rho_\uparrow(x, t)}{\partial x} + 2r_{\rm c}(1 - N_\uparrow(t))^\alpha\rho_\downarrow(x, t), \ x_\uparrow \leq x,\\
\frac{\partial\rho_\downarrow(x,t)}{\partial t} = - v\frac{\partial \rho_\downarrow(x, t)}{\partial x} + 2r_{\rm c}(N_\uparrow(t))^\alpha\rho_\uparrow(x, t), \ x \leq x_\downarrow .
\end{cases}
\end{eqnarray}

\section{Consumption in the Steady State}
\label{AppB}
\begin{figure}[h]
    \centering
    \includegraphics[width=1.0\textwidth]{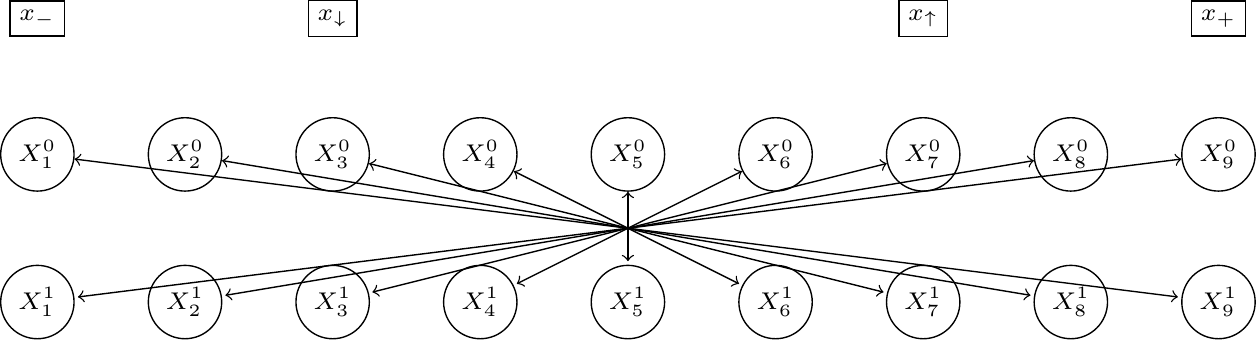}
    \caption{Graphical representation of $T_{\sigma\sigma'}$. Transition probabilities are all equal to $1$.}
    \label{fig13}
\end{figure}

Assume that the steady-state of the master equation Eq.~\eqref{eq:master} is $\rho^{(\rm st)}$.  Then, it satisfies $$\sum_{\sigma'}p_{\sigma\sigma'}\left(\rho^{(\rm st)}\right)\rho^{(\rm st)}_{\sigma'}=\rho^{(\rm st)}_\sigma.$$ We aim to show that $N_{\uparrow}^{(\rm st)} = \sum_\sigma U_\sigma \rho_\sigma^{(\rm st)}=\frac{1}{2}$. In order to prove it, let us consider the linear transformation $T_{\sigma\sigma'}$ which acts on the state $\rho'_\sigma = \sum_{\sigma'}T_{\sigma\sigma'}\rho_\sigma$ and makes the following changes: swaps on and off and reverse order of $X$. $T_{\sigma\sigma'}$ is also a stochastic matrix Fig.~\ref{fig13}. Other important properties of the matrix $T$ are: $T^2 = \mathbbm{1}$ and $TPT = P$, where $P$ is the transition matrix in the case without the mean field control, $\mathbbm{1}$ is the identity matrix. $TPT = P$ directly follows from the symmetry of the $P$ matrix with respect to such a transformation $T$. This two properties lead to the following commutation relation:

\begin{equation}
TP=PT
\end{equation}

Using this commutation relation one derives
\begin{eqnarray}
&&P\rho = \rho\nonumber\\
&&TP\rho = T\rho\nonumber\\
&&PT\rho = T\rho
\end{eqnarray}

In the case when we have only one steady state, $T\rho = \rho$. The relation is satisfied only if $\sum_\sigma U_\sigma\rho_\sigma = 1/2$. Therefore, this property is a consequence of the transition matrix symmetry. (Notice that we do not consider here a more complicated case of multiple competing steady states.) 

\section{Oscillating dynamics}
\label{AppC}

\subsection{Variational principle}
In order to simplify the original master equation one can use a variational principle. to search for approximate solution within a defined class of functions. 

Let us start with the master equation
\begin{eqnarray}
\sum_{\sigma'}p_{\sigma\sigma'}(\rho(t))\rho_{\sigma'}(t) = \rho_{\sigma}(t+1).
\end{eqnarray}
and consider the following functional
\begin{eqnarray}\label{functional}
L(\rho^{L}, \rho^{R})=\sum_{\sigma,\sigma',t}\rho^{L}_{\sigma}(t)p_{\sigma\sigma'}(\rho^{R}(t))\rho_{\sigma'}^{R}(t)-\sum_{\sigma, t}\rho^{L}_\sigma(t)\rho^{R}_\sigma(t+1),
\end{eqnarray}
where $\rho^R$ is a probability mass function and $\rho^L$ is an auxiliary vector (conjugated distribution). Observe that the stationary point of this functional results in the master equation
\begin{eqnarray}
0 = \frac{\partial L(\rho^{L}, \rho^{R})}{\partial \rho^{L}_k(t)} = \sum_{\sigma'}p_{k\sigma'}(\rho^{R}(t))\rho^{R}_{\sigma'}(t) - \rho^{R}_{k}(t+1).
\end{eqnarray}
Using this variational principle one can explore different class of functions and try to find the best solution from this family by minimizing of the functional Eq.~\eqref{functional}.

\subsection{Theoretical explanation of instability}

Let us use the variational formulation to gain a qualitative explanation of the discretization-related instability discussed in the main part of the paper. We derive
\begin{eqnarray}
\rho^{L}(t) = \begin{bmatrix}v_{\uparrow}^{L}(t)\\\vdots\\v_{\uparrow}^{L}(t)\\v_{\downarrow}^{L}(t)\\\vdots\\v_{\downarrow}^{L}(t)
\end{bmatrix}, \ \ \rho^{R}(t) = \begin{bmatrix}v_{\uparrow}^{R}(t)\\\vdots\\v_{\uparrow}^{R}(t)\\v_{\downarrow}^{R}(t)\\\vdots\\v_{\downarrow}^{R}(t)
\end{bmatrix} = 
\begin{bmatrix}\frac{N_{\uparrow}(t)}{n}\\\vdots\\\frac{N_{\uparrow}(t)}{n}\\\frac{(1 - N_{\uparrow}(t))}{n}\\\vdots\\\frac{(1 - N_{\uparrow}(t))}{n}
\end{bmatrix}
\end{eqnarray}
where $v_{\downarrow}^{R}(t)$, $v_{\downarrow}^{L}(t)$, $v_{\uparrow}^{R}(t)$, $v_{\uparrow}^{L}(t)$ are new variables. This type of variational ansatz enforces uniform distribution for ON and OFF states along coordinate (temperature) form $x_-$ to $x_+$. Changing variables, $nv_{\downarrow}^{R}(t) = N_{\downarrow}(t)$, $nv_{\uparrow}^{R}(t) = N_{\uparrow}(t)$, where $n$ is total number of states, results in the following system of equations
\begin{eqnarray}
\small{
\begin{cases}
N_{\uparrow}(t+\Delta t) = \left[\frac{n-1}{n} - \frac{n_{\rm out}/2-1}{n}f\left(r\left(2N_{\uparrow}(t)\right)^\alpha\right)\right]N_{\uparrow}(t) + \left[\frac{1}{n} + \frac{n_{\rm out}/2-1}{n}f\left(r\left(2N_{\downarrow}(t)\right)^\alpha\right)\right]N_{\downarrow}(t), \\
N_{\downarrow}(t+\Delta t) = \left[\frac{1}{n} + \frac{n_{\rm out}/2-1}{n}f\left(r\left(2N_{\uparrow}(t)\right)^\alpha\right)\right]N_{\uparrow}(t) + \left[\frac{n - 1}{n} - \frac{n_{\rm out}/2-1}{n}f\left(r\left(2N_{\downarrow}(t)\right)^\alpha\right)\right]N_{\downarrow}(t),
\end{cases}
}
\end{eqnarray}
where $n_{\rm out}$ is the number of states which are outside  of the comfort zone. Using normalization condition, $N_{\downarrow} + N_{\uparrow} = 1$, we reduce the system to a single equation
\begin{eqnarray}\label{dynamicEq}
&&N_{\uparrow}(t+\Delta t) = \left[\frac{n-2}{n} - \frac{n_{\rm out}/2-1}{n}\left(f(r(2N_\uparrow(t))^\alpha) + f(r(2(1-N_\uparrow(t)))^\alpha)\right)\right]N_{\uparrow}(t)\nonumber\\&& + \left[\frac{1}{n} + \frac{n_{\rm out}/2 - 1}{n}f(r(2(1 - N_{\uparrow}(t)))^\alpha)\right].
\end{eqnarray}

Consider a small perturbation around the stationary state: $N_{\uparrow}(t) = \frac{1}{2} + \delta N(t)$. Linearized version of Eq.~(\ref{dynamicEq}) becomes
\begin{eqnarray}\label{instability}
\delta N(t+\Delta t) = \delta N(t) \left[\frac{n-2}{n} - 2\frac{n_{\rm out}/2-1}{n}(1+\alpha)r\right]
\end{eqnarray}
Analysis of this relation shows emergence of the 3 distinct regimes in the space-time-quantized model which are interpreted as follows
\begin{itemize}
    \item $0\leq\frac{n-2}{n} - 2\frac{n_{\rm out}/2-1}{n}(1+\alpha)r\leq1$, mean field control speeds up relaxation,
    \item $-1\leq\frac{n-2}{n} - 2\frac{n_{\rm out}/2-1}{n}(1+\alpha)r\leq0$, mean field control is too strong and it leads to the perturbation alternating its sign at every step, while the absolute value of perturbation is still decaying,
    \item $\frac{n-2}{n} - 2\frac{n_{\rm out}/2-1}{n}(1+\alpha)r\leq -1$, mean field control changes sign and increases absolute value of perturbation, resulting in the instability.
\end{itemize}

Let us now make a brief comment on the absence of discretization instability in the space-time-continuous model. Consider Eq.~\eqref{instability} in the continuous case:
\begin{eqnarray}
\delta N(t + dt) = \delta N(t)\left[1 - \Gamma(1 + \alpha)r\right],\ \Gamma = \frac{n_{\rm out}}{n}.
\end{eqnarray}
In this case $r\rightarrow 0$, $dt \rightarrow 0$ and $\frac{r}{dt} = {\rm constant}$, so that the factor next to $\delta N(t)$ in the equation never becomes negative. As a result the continuous dynamics is never unstable. The  dynamics never becomes unstable.

Notice that the instability occurs in simulations when one uses a too large time step. This effect is associated with the fact that in the discrete time we may have $1-\Gamma(1 + \alpha)$ either negative or positive, and as $\delta N$ is not continuous, the difference $(\delta N(t+dt)-\delta N(t))$ (slope) is always pointing towards the axis because $(\delta N(t+dt)-\delta N(t))/\delta N(t)<0$. Note that this observation also applies to the continuous time models, however since $\delta N(t)$ varies continuously it only results in decay (towards zero). The comparison between the numerical analysis and the simple theoretical estimation discussed above is shown in Fig.~\ref{fig10}.

\section{Some properties of the spectrum of $S$}
\label{AppD}
Let us assume that the spectrum $\{\Lambda^{(i)}\}$ and the set of right eigenvectors $\{\psi_\sigma^{(i)}\}$ are known. Then we write 
\begin{equation}
    \sum_{\sigma'} S_{\sigma\sigma'}\psi_{\sigma'}^{(i)} = \Lambda^{(i)}\psi_{\sigma}^{(i)}.
\end{equation}
Since,  $\sum_{\sigma} S_{\sigma\sigma'}=1$, one also derives
\begin{equation}
    \Psi^{(i)} = \Lambda^{(i)}\Psi^{(i)},
    \label{appD-central-eq}
\end{equation}
where $\Psi^{(i)} = \sum_\sigma \psi_\sigma^{(i)}$. The equality (\ref{appD-central-eq}) implies that there are two types of eigenmodes. The first type is associated with $\Lambda^{(i)}$ being arbitrary and $\sum_\sigma \psi_\sigma^{(i)} = 0$. The second type occurs when $\Lambda^{(i)} = 1$ and as a result, $\sum_\sigma \psi_\sigma^{(i)}$, is not constrained. There is at least one mode of the second type with the following property of the corresponding right eigenvector: $\sum_\sigma\psi_\sigma \neq 0$, since it is otherwise impossible to decompose a vector $\xi_\sigma$ (for which $\sum_{\sigma}\xi_\sigma\neq 0$) as a linear combination of right eigenvectors $\{\psi^{(i)}_\sigma\}$.
Therefore, one reaches the following conclusions about the spectrum of $S$:
\begin{enumerate}
    \item there exists at least one mode with $\Lambda = 1$ and $\sum_\sigma\psi_\sigma \neq 0$;
    \item for those modes that have $\Lambda\neq 1$, the corresponding right eigenvectors satisfy: $\sum_\sigma \psi_\sigma = 0$.
\end{enumerate}


\begin{thebibliography}{10} 

\bibitem{OConnell2014} N. O'Connell, P. Pinson, H. Madsen, and M. O'Malley, \emph{Benefits and challenges of electrical demand response: A critical review}, Renewable and Sustainable Energy Reviews {\bf 39}, 686-699 (2014). 
doi:10.1016/j.rser.2014.07.098

\bibitem{Hammons2008} T. J. Hammons, \emph{Integrating renewable energy sources into European grids}, International Journal of Electrical Power \& Energy Systems {\bf 30}, 462--475 (2008). 
doi:10.1016/j.ijepes.2008.04.010

\bibitem{Critz2013} D. K. Critz, S. Busche, and S. Connors, \emph{Power systems balancing with high penetration renewables: The potential of demand response in Hawaii}, Energy Conversion and Management {\bf 76}, 609--619 (2013).
doi:10.1016/j.enconman.2013.07.056

\bibitem{Auer2016} H. Auer and R. Haas, \emph{On integrating large shares of variable renewables into the electricity system}, Energy {\bf 115}, 1592--1601 (2016).
doi:10.1016/j.energy.2016.05.067  

\bibitem{Albadi2007} M. H. Albadi and E. F.  El-Saadany, \emph{Demand response in electricity markets: An overview}, Proceedings of the 2007 IEEE Power Engineering Society General Meeting,  24-28 June 2007. 
doi:10.1109/PES.2007.385728

\bibitem{Lampropoulos2013} I. Lampropoulos, W. L. Kling, P. F. Ribeiro, and J. van den Berg, \emph{History of demand side management and classification of demand response control schemes}, Proceedings of the 2013 IEEE Power Energy Society General Meeting, pages 1-5. 
doi:10.1109/PESMG.2013.6672715

\bibitem{Lynch2019} M. \'A. Lynch, S. Nolan, M. T. Devine, M. O'Malley, \emph{The impacts of demand response participation in capacity markets}, Applied Energy {\bf 250}, 444--451 (2019).
doi:10.1016/j.apenergy.2019.05.063

\bibitem{Wohlfarth2020} K. Wohlfarth, M. Klobasa, R. Gutknecht, \emph{Demand response in the service sector – Theoretical, technical and practical potentials}, Applied Energy {\bf 258}, 114089 (2020).
doi:10.1016/j.apenergy.2019.114089

\bibitem{Schiel2017} C. Schiel, P. G. Lind, and P. Maass, \emph{Resilience of electricity grids against transmission line overloads under wind power injection at different nodes}, Scientific Reports {\bf 7}, 11562 (2017). 
doi:10.1038/s41598-017-11465-w

\bibitem{Gajduk2014} A. Gajduk, M. Todorovski, J. Kurths and L. Kocarev, \emph{Improving power grid transient stability by plug-in electric vehicles}, New Journal of Physics {\bf 16}, 115011 (2014).

\bibitem{Yesilbudak2018} M. Yesilbudak and A. Colak, \emph{Integration challenges and solutions for renewable energy sources, electric vehicles and demand-side initiatives in smart grids}, 7th International IEEE Conference on Renewable Energy Research and Applications, ICRERA 2018 8567004, pp. 1407-1412. 
doi:10.1109/ICRERA.2018.8567004

\bibitem{Motalleb2016} M. Motalleb, M. Thornton, E. Reihan, and R. Ghorbani, \emph{Providing frequency regulation reserve services using demand response scheduling}, Energy Conversion and Management {\bf 124}, 439--452 (2016).
doi:10.1016/j.enconman.2016.07.049

\bibitem{Chong1979} C.-Y. Chong and A. S. Debs, \emph{Statistical synthesis of power system functional load models}, Proceedings of the 1979 18th IEEE Conference on Decision and Control including the Symposium on Adaptive Processes {\bf 2}, 264--269 (1979). 
doi:10.1109/CDC.1979.270177

\bibitem{Ihara1981} S. Ihara and F. Schweppe, \emph{Physically based modeling of cold load pickup}, IEEE Transactions on Power Apparatus and Systems {\bf 100}, 4142--4150 (1981). 
doi:10.1109/TPAS.1981.316965

\bibitem{Chong1984} C.-Y. Chong and R. P. Malham\'e, \emph{Statistical synthesis of physically based load models with applications to cold load pickup}, IEEE Transactions on Power Apparatus and Systems {\bf 103}, 1621--1628 (1984).
doi: 10.1109/TPAS.1984.318643

\bibitem{Gardiner2004} C. W. Gardiner, \emph{Handbook of stochastic methods for physics, chemistry and the natural sciences, 3rd ed.} (Springer Series in Synergetics, vol. 13, Berlin: Springer-Verlag, 2004).

\bibitem{VanKampen2007} N. van Kampen, \emph{Stochastic Processes in Physics and Chemistry (Third Edition)} (Amsterdam: Elsevier, 2007).

\bibitem{Malhame1985} R. Malham\'e, R. and C.-Y. Chong, \emph{Electric load model synthesis by diffusion approximation of a high-order hybrid-state stochastic system}, IEEE Transactions on Automatic Control {\bf 30}, 854--860 (1985).
doi:10.1109/TAC.1985.1104071

\bibitem{Malhame1988} R. Malham\'e and C.-Y. Chong, \emph{On the statistical properties of a cyclic diffusion process arising in the modeling of thermostat-controlled electric power system loads}, SIAM Journal on Applied Mathematics {\bf 48}, 465--480 (1988).
doi:10.1137/0148026

\bibitem{ElFerik1994} S. El-Ferik and R. P. Malham\'e, \emph{Identification of alternating renewal electric load models from energy measurements}, IEEE Transactions on Automatic Control {\bf 39}, 1184--1196 (1994).
doi:10.1109/9.293178

\bibitem{Lu2004} N. Lu and D. Chassin, \emph{A state-queueing model of thermostatically controlled appliances}, IEEE Transactions on Power Systems {\bf 19}, 1666--1673 (2004).
doi:10.1109/TPWRS.2004.831700

\bibitem{Lu2005} N. Lu, D. Chassin, and S. Widergren, \emph{Modeling uncertainties in aggregated thermostatically controlled loads using a state queueing model}, IEEE Transactions on Power Systems {\bf 20}, 725--733 (2005).
doi:10.1109/TPWRS.2005.846072

\bibitem{Callaway2009} D. S. Callaway, \emph{Tapping the energy storage potential in electric loads to deliver load following and regulation, with application to wind energy} Energy Conversion and Management {\bf 50}, 1389-1400 (2009).
doi:10.1016/j.enconman.2008.12.012

\bibitem{Callaway2011} D. S. Callaway and I. A. Hiskens,  \emph{Achieving controllability of electric loads} Proceedings of the IEEE {\bf 99}, 184–199 (2011).
doi:10.1109/JPROC.2010.2081652

\bibitem{Bashash2011} S. Bashash and H. K. Fathy, \emph{Modeling and control insights into demand-side energy management through setpoint control of thermostatic loads}, Proceedings of the 2011 American Control Conference, 4546–4553 (2011).
doi:10.1109/ACC.2011.5990939

\bibitem{Perfumo2012} C. Perfumo, E. Kofman, J. H. Braslavsky, and J. K.Ward, \emph{Load management: Model-based control of aggregate power for populations of thermostatically controlled loads}, Energy Conversion and Management {\bf 55}, 36--48 (2012).
doi:10.1016/j.enconman.2011.10.019

\bibitem{Chassin2015} D. P. Chassin, J. Stoustrup, P. Agathoklis, and N. Djilali, \emph{A new thermostat for real-time price demand response: Cost, comfort and energy impacts of discrete-time control without deadband}, Applied Energy {\bf 155}, 816--825 (2015).
doi:10.1016/j.apenergy.2015.06.048

\bibitem{Wei2018} C. Wei, J. Xu, S. Liao, Y. Sun, Y. Jiang, and Z. Zhang, \emph{Coordination optimization of multiple thermostatically controlled load groups in distribution network with renewable energy}, Applied Energy {\bf 231}, 456--467 (2018).
doi:10.1016/j.apenergy.2018.09.105

\bibitem{Chertkov2017} M. Chertkov and V. Chernyak, \emph{Ensemble of thermostatically controlled loads: Statistical physics approach}, Scientific Reports {\bf 7}, 8673 (2017). 
doi:10.1038/s41598-017-07462-8

\bibitem{Metivier2019} D. M\'etivier, I. Luchnikov, and M. Chertkov, \emph{Power of ensemble diversity and randomization for energy aggregation}, Scientific Reports {\bf 9}, 5910 (2019).
doi:10.1038/s41598-019-41515-4

\bibitem{Metivier2020} D. M\'etivier and M. Chertkov, \emph{Mean-field control for efficient mixing of energy loads}, Physical Review E {\bf 101}, 022115 (2020).
doi:10.1103/PhysRevE.101.022115

\bibitem{Rajabi2017} A. Rajabi, L. Li, J. Zhang, and J. Zhu, \emph{Aggregation of small loads for demand response programs -- Implementation and challenges: A review}, 2017 IEEE International Conference on Environment and Electrical Engineering and 2017 IEEE Industrial and Commercial Power Systems Europe (EEEIC / I\&CPS Europe). 
doi:10.1109/EEEIC.2017.7977631

\bibitem{2003HCM} M. Huang, P.E. Caines, R.P. Malhame, \emph{Individual and mass behaviour in large population stochastic wireless power control problems: Centralized and Nash equilibrium solutions}, Proceedings of the 42nd IEEE Conference on Decision and Control (CDC), (2003).
doi:10.1109/CDC.2003.1272542

\bibitem{2005WBV} G.Y. Weintraub, C.L. Benkard, B. Van Roy, \emph{Oblivious Equilibrium: A Mean Field}, Advances in Neural Information Processing Systems, MIT Press, (2005).

\bibitem{2006LL} J.-M. Lasry, P.-L. Lions, \emph{Jeux a champ Moyens I,II}, Comptes Rendus Math\'ematiques {\bf 343}, 619--625 (2006). 
doi:10.1016/j.crma.2006.09.019

\bibitem{2006LLb} J.-M. Lasry, P.-L. Lions,  \emph{Mean Field Games}, Japanese Journal of Mathematics, {\bf 2}, 229--260 (2007).
doi:10.1007/s11537-007-0657-8

\bibitem{2006HMC} M. Huang, R.P. Malham\'e, P.E. Caines, \emph{Large Population stochastic dynamic games: Closed loop McKean-Vlasov  and the Nash certainty equivalence principle}, Communications in Information \& Systems {\bf 6},221--252 (2006).
https://projecteuclid.org/euclid.cis/1183728987

\bibitem{Bensoussan2013} A. Bensoussan, J. Frehse, and P. Yam, \emph{Mean Field Games and Mean Field Type Control Theory} (Springer-Verlag New York, 2013).

\bibitem{ashrae} ANSI/ASHRAE Standard 55-2013: Thermal environmental conditions forhuman occupancy.

\bibitem{Fanger1970} P. O. Fanger, \emph{Thermal comfort. Analysis and applications in environmental engineering} (Danish Technical Press, Copenhagen, 1970).

\bibitem{Satish2012} U. Satish, M. J. Mendell, K. Shekhar, T. Hotchi, D. Sullivan, S. Streufert, et al. \emph{Is CO$_2$ an indoor pollutant? Direct effects of low-to-moderate CO$_2$ concentrations on human decision-making performance}, Environmental Health Perspectives {\bf 120}, 1671--1677 (2012).

\bibitem{Allen2016} J. G. Allen, P. Mac Naughton, U. Satish, S. Santanam, J. Vallarino, J. D. Spengler, \emph{Associations of cognitive function scores with carbon dioxide,ventilation, and volatile organic compound exposures in office workers: a controlled exposure study of green and conventional office environments}, Environmental Health Perspectives {\bf 124}, 805--812 (2016). 

\bibitem{Sinitsyn2012} N. A. Sinitsyn, S. Kundu, and S. Backhaus, \emph{Safe protocols for generating power pulses with heterogeneous populations of thermostatically controlled loads}, Energy Conversion and Management {\bf 67}, 297--308 (2012).
doi:10.1016/j.enconman.2012.11.021

\bibitem{Mehta2014} N. Mehta, N. A. Sinitsyn, S. Backhaus, B. C. Lesieutre, \emph{Safe control of thermostatically controlled loads with installed timers for demand side management}, Energy Conversion and Management {\bf 86}, 784--791 (2014).
doi:10.1016/j.enconman.2014.06.049

\bibitem{Ryzhov2019} A. Ryzhov, H. Ouerdane, E. Gryazina, A. Bischi, and K. Turitsyn, \emph{Model predictive control of indoor microclimate: existing building stock comfort improvement}, Energy Conversion and Management {\bf 179}, 219--228  (2019).
doi:10.1016/j.enconman.2018.10.046

\bibitem{Liu2020} Y. Liu, M. S. Mauter, \emph{Assessing the demand response capacity of U.S. drinking water treatment plants}, Applied Energy {\bf 267}, 114899 (2020).
doi:10.1016/j.apenergy.2020.114899

\bibitem{Hassan2020} A. Hassan, S. Acharya, M. Chertkov, D. Deka and Y. Dvorkin, \emph{A hierarchical approach to multi-energy demand response: From electricity to multi-energy applications}, Special Issue on Multi-Energy Systems, Proceedings of IEEE {\bf 108}, 1457--1474 (2020). 
doi:10.1109/JPROC.2020.2983388


\end{thebibliography}
\end{document}